\documentclass[twocolumn]{aastex63}

\usepackage{hyperref}
\usepackage{amsmath}
\usepackage{rotating}
\usepackage{float}

\accepted{16.06.2022}

\submitjournal{ApJ}

\shorttitle{The Effects of Cosmic Rays on the Chemistry of Dense Cores}
\shortauthors{O'Donoghue et al.}

\begin{document}

\title{The Effects of Cosmic Rays on the Chemistry of Dense Cores}

\correspondingauthor{Ross O'Donoghue}
\email{ross.o'donoghue.19@ucl.ac.uk}

\author{Ross O'Donoghue}
\affiliation{Department of Physics \& Astronomy,
University College London,
Gower Street,
London,
WC1E 6BT,
UK}

\author{Serena Viti}
\affiliation{Leiden Observatory,
Leiden University,
PO Box 9513,
2300 RA Leiden,
The Netherlands}

\affiliation{Department of Physics \& Astronomy,
University College London,
Gower Street,
London,
WC1E 6BT,
UK}

\author{Marco Padovani}
\affiliation{INAF-Osservatorio Astrofisico di Arcetri - Largo E. Fermi, 5 - 50125 Firenze, Italy}

\author{Tomas James}
\affiliation{Department of Physics \& Astronomy,
University College London,
Gower Street,
London,
WC1E 6BT,
UK}

\begin{abstract}
    Cosmic rays are crucial for the chemistry of molecular clouds and their evolution. They provide essential ionizations, dissociations, heating and energy to the cold, dense cores. As cosmic rays pierce through the clouds they are attenuated and lose energy, which leads to a dependency on the column density of a system. The detailed effects these particles have on the central regions still needs to be fully understood. Here, we revisit how cosmic rays are treated in the UCLCHEM chemical modeling code by including both ionization rate and H$_2$ dissociation rate dependencies alongside the production of cosmic ray induced excited species and we study in detail the effects of these treatments on the chemistry of pre-stellar cores. We find that these treatments can have significant effects on chemical abundances, up to several orders of magnitude, depending on physical conditions. The ionization dependency is the most significant treatment, influencing chemical abundances through increased presence of ionized species, grain desorptions and enhanced chemical reactions. Comparisons to chemical abundances derived from observations show the new treatments reproduce these observations better than the standard handling. It is clear that more advanced treatments of cosmic rays are essential to chemical models and that including this type of dependency provides more accurate chemical representations.

\end{abstract}


\section{Introduction} \label{sec:intro}
    Cosmic rays (CR) play a vital role in the chemistry of cold (10-30K), dense ($>$ 10$^2$ cm$^{-3}$) molecular clouds as they can pierce deep into them, unlike the interstellar UV radiation \citep[for a review see][]{Indriolo2013}. These high energy interstellar particles, primarily consisting of protons but can be heavier elements and electrons, have large energy ranges, up to ZeV energies \citep{BLANDFORD2014_CROrigins}. Although the energies can be high, it is the lower energy CRs ($\leq$ 1 TeV) that affect the dense interiors \citep{CosmicRaysintheInterstellarMedium_CPT2,pado_CR_review}. In these regions, CRs have a wide variety of effects, one of the most important being a producer of atomic hydrogen through the dissociation of H$_2$ \citep{vanderWerf_88_high-res-L134, montgomery_1995_HI-study-riegel-coldcloud, li2003h1_self_adsorb, golsmith2005_HISA-darkcloud, Padovani_2018-diss}. Other important effects are: being the dominant source of ionization; regulating the degree of coupling of the gas and the magnetic field;  having an important role on the dynamics and the collapse timescale of collapsing clouds \citep[e.g.,][]{padovani_collapsing_cores...560A.114P, padovani_magnetic_field...571A..33P}; providing heating and energy to dust grains \citep{1973_Grain_Heating,shingledecker2018b,WGH_Kalvans_2019,Sipila2020,Sipila2021,Silsbee2021}; producing internal UV photons   \citep{CR_UV_Generation1983}; may have a role on the charge distribution on dust grains \citep{IS_dust_charging2015}; influencing disk growth \citep{Ionization_disk_size_2020}; and affecting deuteration \citep{2008CRDeuteration}. For example, each species ionised by a CR releases an electron. This secondary electron can cause further collisions, which in turn, depending on the energy, can induce more ionizations and heating \citep{Ivlev_2021...909..107I}. If a secondary electron does not have enough energy to ionise a species, the species may become excited \citep{shingledecker2018a}. Excited species produced by CR bombardment have energy levels higher than their base counterparts, allowing these excited species to overcome some reaction barriers that would otherwise be difficult in cold environments. These species have been shown to drive more complex chemistry from reactions that can form interstellar complex organic molecules \citep{Abplanalp2016}.
    \\
    
    Although CRs can pierce deep into the molecular clouds, they are still attenuated as they collide and lose energy. The  denser the region is, the lower the CR ionization rate becomes \citep{Padovani_2018-ion}. This leads to a dependency of the ionization rate on the density of a region, more precisely on the H$_2$ column density passed through by CRs.
    \\

    As the Earth is shielded from the low-energy spectrum of CRs through solar modulation \citep[see][for a review on solar modulation]{Potgieter2013SolarMOReview}, measurements of the CR ionization rate taken from Earth are not indicative of measurements in the interstellar medium (ISM) and are in fact lower. Observations of molecules that are dependant or sensitive to the CR ionization rate (for example, H${_3^+}$ is produced from CR ionization of H$_2$) can be used as a tracer for the ionization rate \citep[see][for a review]{CosmicRaysintheInterstellarMedium_CPT2}.
    The "typical" value for the CR ionization rate is often taken to be around the order of $10^{-17}$s$^{-1}$ \citep[e.g.,][]{Spitzer_Tomasko1968ApJ...152..971S,Solomen_Werner1971ApJ...165...41S,Herbst_Kemp1973ApJ...185..505H,li2003h1_self_adsorb}. It is necessary to note that while this may be known as the "typical" rate, observations shown environments with significantly higher rates. Diffuse clouds have been observed with ionization rates in the order of 10$^{-16}$s$^{-1}$ \citep{Indriolo2007_h3+diffuse,Indriolo_2012_H3+} and rates of up to 10$^{-14}$s$^{-1}$ have been observed within the inner 300 parsecs of the Galactic centre \citep{Oka2006-galactic,Petit-2016-galactic}.
    Recently both Voyager spacecrafts have passed beyond the heliopause, and have been observing lower energies of the CR spectrum (as low as 3 MeV for both nuclei and electrons) \citep{Voyager1Data,2019_voyager_2...3.1013S}. This data from the Voyager probes can be used to estimate a lower boundary for the ISM ionization rate \citep{IS_dust_charging2015, Padovani_2018-ion}. In fact, the local CR flux measured by the Voyager probes is thought to be unmodulated by the solar wind. However, the magnetometers on board the Voyager spacecrafts have not yet detected a change in the magnetic field direction, as would be expected if they had passed the heliopause \citep{gloecker_2015_voyager}. Furthermore, the ionization rate using the fluxes from Voyager only give a lower limit to the observational estimates in nearby diffuse molecular clouds \citep[e.g.,][]{Indriolo_2012_H3+}. 
    \\
    
    The hydrogen chemistry of CRs is essential to the chemical evolution of a cloud. H$_3^+$ is fundamental to the production of many polyatomic gas-phase molecules \citep{Herbst_Kemp1973ApJ...185..505H} and is formed through the CR ionization reaction: 
    \begin{equation*}
        {\rm H_2+CR}\rightarrow{\rm H_2^++ } e^-\rm{+CR^\prime}
    \end{equation*}
    and the subsequent reaction
    \begin{equation*}
        {\rm H_2^++H_2}\rightarrow{\rm H_3^++H}
    \end{equation*}    
    where H$_2$ ionization is the rate limiting reaction. In dense clouds H$_3^+$ can then react via proton transfer with molecules such as CO (to form HCO$^+$ or HOC$^+$), O (forming OH$^+$), N$_2$ (forming HN$_2^+$) and HD (forming H$_2$D$^+$). See the review by \textcite{Indriolo2013} for a more in-depth summary.\\
    
    CRs also dissociate molecular hydrogen in the ISM through the reaction:
    \begin{equation*}
        {\rm H_2+CR}\rightarrow{\rm H+H+CR^\prime}
    \end{equation*}
    In high density regions this reaction is the only form of H$_2$ dissociation, as the UV photons for photodissociation cannot penetrate deep into the cloud. This reaction depends on the CR dissociation rate, which is often taken to be equal to the ionization rate. In chemical networks however, the rate is often lower than the "typical" value. In UMIST \citep{UMIST12McElroy_2013} for example, the  H$_2$ dissociation rate is 1.30$\times$10$^{-18}$ s$^{-1}$. In \textcite{Padovani_2018-diss}, it has been shown that the H$_2$ dissociation rate is higher than is often represented in chemical networks, is not a constant value and is not equal to the ionization rate. The rate is dependent on the secondary electrons produced from CR ionization and can be represented as a function of column density, similar to the CR ionization rate in Equation \ref{equation:polynomialfit} \citep{Padovani_2018-diss}.
    \\
    
    As discussed (and as seen in similar works like \textcite{Redaelli2021}) it is clear that CRs are extremely important to the chemistry of molecular clouds and their evolution, and hence it is essential that their effects are represented accurately within modern chemical models. This paper aims to improve the handling of CRs in gas-grain chemical models, by introducing both the CR ionization rate and the H$_2$ dissociation rate as functions of column density and to include the ability to produce excited species and their reactions on the grain. The chemical effects of these additions will be tested on models of collapsing cloud cores. These environments are crucial steps in the early stages of star formation and the effects of CRs on these objects where the gas density increases with time and changes the column density of the core, still need to be investigated. In section \ref{sec:modelling} we discuss the chemical modelling and detail the CR treatments we have included for this paper. In section \ref{results} we describe the effects these treatments have on the chemical abundances of selected species and discuss the main processes involved in these changes, while in section \ref{summary} we summarize our findings.

\section{Modelling} \label{sec:modelling}
    
    The chemical code selected for this paper is UCLCHEM \citep{Holdship2017UCLCHEMRELEASE}. UCLCHEM is a time dependent gas-grain chemical code, written in modern Fortran. UCLCHEM is an open sourced chemical code, freely available for use and modification. It is diverse in use due to its modular nature. Specific environments (shocks, cores, collapses) each have their own physics module. UCLCHEM uses separate gas and grain networks. The default gas phase network used is the UMIST RATE12\footnote{\url{www.udfa.net}} network, described in \textcite{UMIST12McElroy_2013} and is used for this paper. The grain network used is described in section \ref{sec:excited_species_additions} below. For more detailed information on UCLCHEM see \cite{Holdship2017UCLCHEMRELEASE} or visit the \href{https://uclchem.github.io/}{UCLCHEM website}.
    \\

    \subsection{Treating CRs in  UCLCHEM} \label{sec:additions_to_uclchem}

        \subsubsection{CR Ionization Rate} \label{sec:cosmic_ray_ionisation_additions}
            
            In \textcite{Padovani_2018-ion} a polynomial fit was developed to express the dependency of the CR ionization rate on column density. We have implemented such a fit into UCLCHEM. 

            \begin{equation} \label{equation:polynomialfit}
                \log_{10} \frac{\zeta}{\rm s^{-1}} = \sum_{k\geq0} c_k  \log^k_{10} \frac{N}{\rm cm^{-2}} \, ,
            \end{equation}
        
            where $k$ is an integer from 0 to 9, $c_k$ is the fitting coefficient and $N$ is the column density.
            \\
            
            Equation \ref{equation:polynomialfit} is used to calculate the ionization rate at each time step and the calculated rate is used in all chemical reactions that involve the CR ionization rate.\\
            
            Table \ref{tab:coefficients} gives two sets of fitting coefficients. One, labelled as model $L$, describes the trend of the ionization rate as a function of the column density obtained by using the Voyager data; the other, labelled as model $H$, represents the average value of the ionization rate in diffuse clouds \citep{Neufeld_2010, Shaw_2008,Indriolo_2012_H3+,Neufeld_2017}. 

        \begin{figure*}
            \centering
            \includegraphics[width = 13cm]{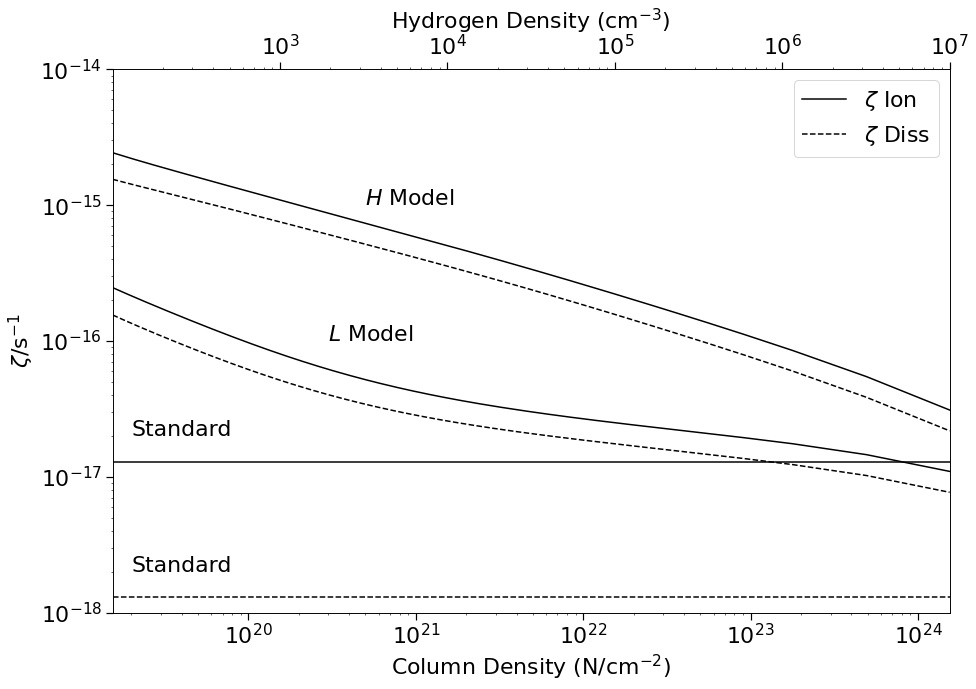}
            \caption{Figure showing the CR ionization rate dependency (solid line) and the H$_2$ dissociation rate dependency (broken line) compared to the standard UCLCHEM handling. For these models we use a cloud size of 0.05 parsecs.}
            \label{fig:ionDependency}
        \end{figure*}

        \subsubsection{\texorpdfstring{H\textsubscript{2} Dissociation Rate}{H2 Dissociation Rate}} \label{sec:cr dissociation additions}
        
            \textcite{Padovani_2018-diss} evaluated the H$_2$ dissociation rate based on the same CR interstellar spectra used to compute the ionization rate. In the following, the dissociation rate is parameterised as a function of the column density by using Equation \ref{equation:polynomialfit} with the coefficients listed in Table \ref{tab:coefficients}. Similarly to the CR ionization treatment, the dissociation rate can now be calculated at each time step of the model. However, this handling of the dissociation rate can only be activated if the CR ionization dependency is also activated and is automatically set to the same model ($L$ or $H$) as the CR ionization dependency.
            
            Figure \ref{fig:ionDependency} shows how the CR ionization rate and H$_2$ dissociation rate differ from the 'standard' handling of UCLCHEM (i.e. the fixed, user-defined value) under increasing density. This particular example shows all three models at a $\times$1 ionization factor. UCLCHEM handles the ionization rate in multiples of 1.3$\times$10$^{-17}$s$^{-1}$, so an ionization factor of $\times$1 will correspond to an ionization rate of 1.3$\times$10$^{-17}$s$^{-1}$ and a H$_2$ dissociation rate of 1.3$\times$10$^{-18}$s$^{-1}$ for the standard handling. UCLCHEM calculates the column density by multiplying the size of the cloud (in parsecs) by the total hydrogen density (cm$^{-3}$).
            
        \subsubsection{Excited Species} \label{sec:excited_species_additions}

            UCLCHEM uses a user defined grain network, separate from the gas phase network. The default grain network that is provided with UCLCHEM handles some basic CR and photon interactions, freeze out reactions and chemical desorption and diffusion reactions. This network was used for this paper, with the additions of excited species production and reactions due to CRs: these excited species are added using the principles described in detail in \textcite{shingledecker2018a} and used in \textcite{shingledecker2018b}. \\
            
            The underlying principles are that CR bombardments of a solid species generally have one of the following outcomes:

            \begin{eqnarray*}
            {\rm (R1)}~{\rm A+CR} &\rightarrow& {\rm A^+} + e^- + {\rm CR^\prime} \\\nonumber
            {\rm (R2)}~{\rm A+CR} &\rightarrow& {\rm B^* + C^* + CR^\prime} \\\nonumber
            {\rm (R3)}~{\rm A+CR} &\rightarrow& {\rm B + C + CR^\prime} \\\nonumber
            {\rm (R4)}~{\rm A+CR} &\rightarrow& {\rm A^* + CR^\prime} \\\nonumber
            \end{eqnarray*}

            Where A is the target species, B and C are dissociated products and $^*$ represents an excited species. The reaction rates for these interactions are defined in \cite{shingledecker2018a} and follow the formula:
            \begin{equation} \label{equation:exReactionRate}
                k_{Rn}=G_{\rm Rn}\left(\frac{S_e}{100~\rm eV}\right) \left(\phi_{\rm ST} \left[\frac{\zeta}{10^{-17} \rm s^{-1}} \right] \right)
            \end{equation}
            
            Where $G_{\rm Rn}$ is the radiochemical yield for the reaction pathway Rn, (Rn being R1 to R4 above), $S_e$ is the electronic stopping cross section. $\phi_{\rm ST}$ is the integrated Spitzer-Tomasko CR flux (from 0.3 MeV to 100 GeV) and has a value of 8.6 cm$^{-2}$ s$^{-1}$. $\zeta$ is a CR ionization rate scaling factor the CR flux.\\
            
            Once produced, an excited species will either react with another solid species, or relax back to the ground energy state. The excited species reaction proceeds at the rate:
            \begin{equation} \label{exReactionRate}
                k_{st} = f_{br}\left[\frac{v_0^B + v_0^A}{N_{site}n_{dust}}\right]
            \end{equation}
            
            Where $f_{br}$ is the branching ratio, $v_0$ is the vibrational frequency of the species, $N_{site}$ is the number of physisorption sites on the grain and $n_{dust}$ is the dust density. \\

    \subsection{Modelling pre-stellar cores}
    
        The effects of these additions will be studied in the cases of pre-stellar cores. Pre-stellar cores represent the early stages of low-mass star formation and have densities in the range of $\sim$10$^4$ - 10$^7$ cm$^{-3}$, depending among other things, on their evolutionary stage. The models will be set to mimic these regions; in each case, the model will start at an initial density of  10$^2$ cm$^{-3}$. At $\sim$10$^6$ years, the models will collapse in free fall to a specific final density. To cover the  density range, the models have four possible final densities: 10$^4$, 10$^5$, 10$^6$ or 10$^7$ cm$^{-3}$. After collapsing, the models are set to run with static conditions until a final time of 10$^8$ years is reached, in order to investigate the chemical evolution over time. To determine the influence that temperature and radiation field may have with the CR ionization dependency, each will be varied independently (see Table \ref{tab:tablesmallparams} for values). Note: UCLCHEM assumes 1 Habing to be the Galactic radiation field strength. The chemical species that will be analysed are: H$_2$O, CS, NH$_3$\textsubscript{(grain)}, N$_2$H$^+$, NH$_3$, CO\textsubscript{(grain)}, HCO$^+$, H$_2$O\textsubscript{(grain)} and CO$_2$\textsubscript{(grain)}. These species are important as some act as tracers of the gas-phase of pre-stellar cores, their regions and physical conditions (CS, N$_2$H$^+$)\citep{Lee_1999}, some are key species for grain chemistry and the chemical complexity (NH$_3$, NH$_3$\textsubscript{(grain)})\citep{Rodgers_2001} and others are some of the most abundant species found in these regions (H$_2$O\textsubscript{(grain)},CO\textsubscript{(grain)},CO$_2$\textsubscript{(grain)})\citep{_berg_2011}.
        \\

        The CR ionization dependency, H$_2$ dissociation dependency and excited species production and reactions will all be tested  individually as well as combined. To test the effects of the CR ionization dependency, under each condition, models will be run with the CR ionization dependence turned off and compared to the same conditions with the $L$ and the $H$ model dependencies activated. The H$_2$ dissociation rate dependency can only be activated when the $L$ or $H$ ionization model is also selected. To test the influence of the H$_2$ dissociation rate, $L$ and $H$ models with the dissociation dependency disabled will be compared to the same model with the dissociation dependency activated. The excited species can be activated without the CR or H$_2$ dissociation rate dependencies. As such their effects will be examined independently of the other additions and then combined together. Table \ref{tab:tablesmallparams} shows the summary of the parameters investigated, their values and their descriptions. In total, a grid of 280 models were run. 
        
        \begin{table*}[tbp]
            \centering
                \begin{tabular}{|c|c|c|}
                    \hline
                    \textbf{Parameter} & \textbf{Value} & \textbf{Description} \\
                    \hline
                    \textbf{Final Density (cm$^{-3}$) } & 10$^4$ / 10$^5$ / 10$^6$ / 10$^7$ & Selects the final density of the model \\
                    \hline
                    \textbf{Initial Temperature (K)} & 10 / 20 / 30 & Selects the gas temperature \\
                    \hline
                    \textbf{Radiation field (Habing)} & 1 / 10 / 100 & Adjusts the local interstellar radiation field \\
                    \hline
                    \textbf{$\zeta$} & $\times$1 / $\times$10 / $\times$100 & Adjusts the CR ionization rate as a \\
                    & & multiplicative of 1.3  x 10$^{-17}$s$^{-1}$ for the\\
                    & & standard UCLCHEM handling or adjust \\
                    & & the ionization dependency if desired \\
                    \hline
                    \textbf{Modified Ionization} & Basic / $L$ model / $H$ model & Selects the "Basic" UCLCHEM handling \\
                     & & (fixed, user defined value) or the updated \\
                     & & ionization rate dependency \\
                    \hline
                    \textbf{Modified Dissociation} & Off / On & Toggles modified H$_2$ dissociation rate \\
                    & & (only active with $L$ or $H$ ionization models) \\
                    \hline
                    \textbf{Excited Species} & Off / On & Toggles inclusion of excited species \\
                    \hline
                \end{tabular}
                \caption{Parameters of the model grid for ionization and H$_2$ dissociation.}
                \label{tab:tablesmallparams}
            \end{table*}

\section{Results}
\label{results}

    When discussing the influence of the CR ionization rate dependency on the chemistry of pre-stellar cores, our simulated core evolution is split into three phases. The pre-collapse phase covers the period of up to $\sim$ 10$^6$ years and represents the period of time leading up to the beginning of the cloud collapse (abundances are examined at a time of 10$^5$ years). This phase of the model has a gas density of 10$^2$ cm$^{-3}$, until the collapse phase where density begins to increase. The cloud collapse phase represents the time in which the cloud is undergoing collapse in free fall and occurs between $\sim$10$^6$ - 6$\times$10$^6$ years, depending on the final density. The density here is increasing over this period from the initial density to the selected final density of the model. The post-collapse represents the period of static density, after the cloud collapses to the designated final density. The post-collapse phase will always have a constant density equal to the selected final density parameter of the individual model. In order to assess trends in our simulations, we set a lower limit for an "observable"  fractional abundances of 10$^{-13}$; below this fractional abundance, changes across the parameter space will be considered irrelevant. Additionally, any changes in abundances that are below a factor of 3 will not be discussed, as these differences are not likely observable.

    \subsection{Density Dependent ionization Rates}
    
        \begin{figure*}[tbp]
            \centering
            \includegraphics[width=18cm, height=12cm]{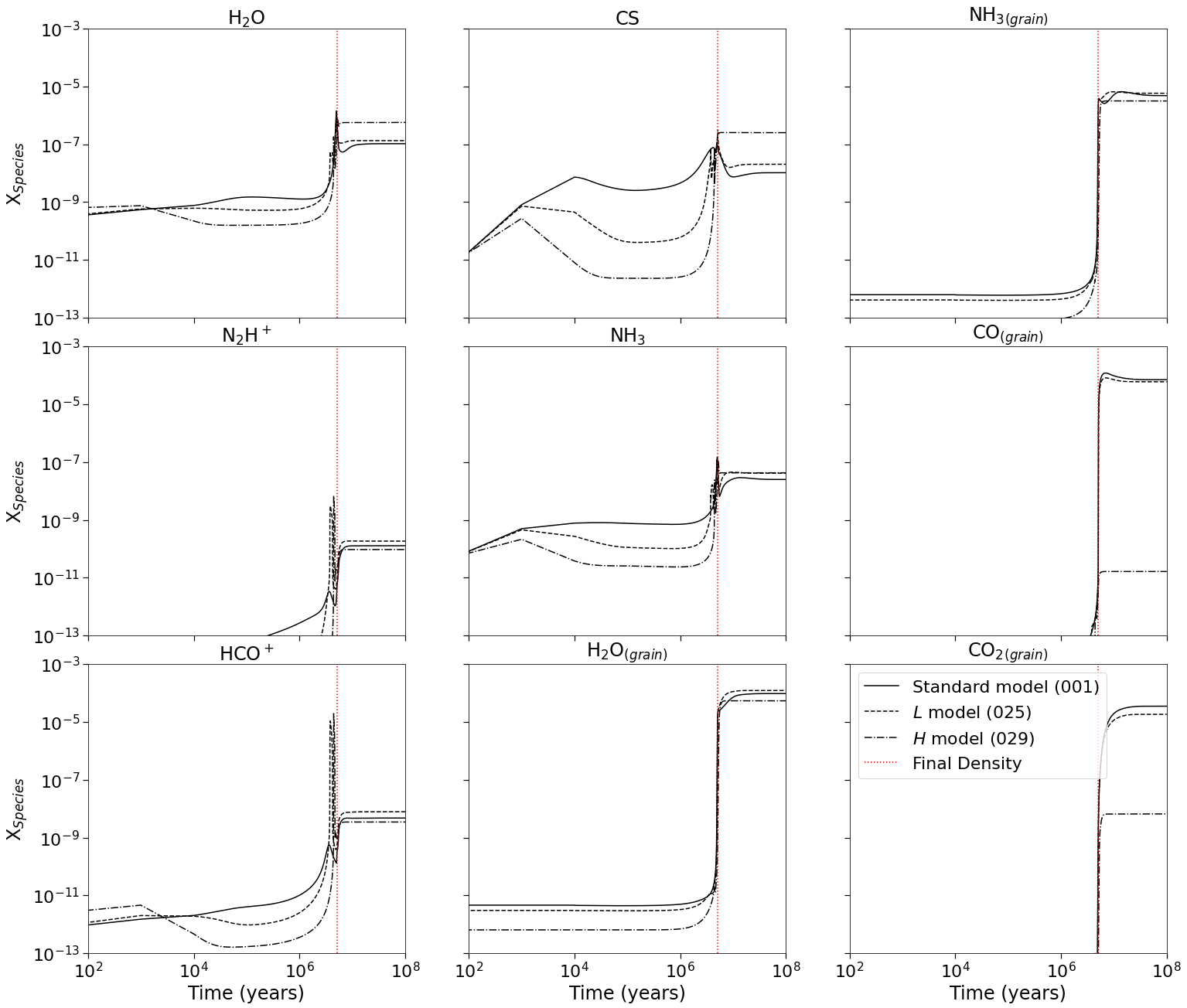}
            \caption{Plot showing the effect of $L$ and $H$ ionization models (short- and long-dashed black lines, respectively) on chemical abundances over time compared to the basic UCLCHEM handling (solid black line). These models have a final density of 10$^4$ cm$^{-3}$, an initial temperature of 10 K, a radiation field of $\times$1 and an ionization rate of $\times$1. The red line represents the time which final density is reached and the numbers in the legend represent the model number identifier.}
            \label{fig:ion_1e4_bas}
        \end{figure*}
    
        Figure \ref{fig:ion_1e4_bas} shows the effects of adding the ionization rate density dependency on chemical abundances for a final density of 10$^4$ cm$^{-3}$, a temperature of 10 K, a radiation field strength of 1 Habing and an un-adjusted ionization rate factor of $\times$1. Table \ref{tab:ionBasResults} summarises the abundance trends. In the pre-collapse phase, the addition of the ionization rate dependency results in reduced abundances for all our selected species. The same trend is seen for both the $L$ and $H$ models with the $H$ model having enhanced effects (i.e. larger reductions in abundances, up to 3 orders of magnitude, see Table \ref{tab:ionBasResults}). We note that during this phase of the pre-stellar core evolution, the CR ionization rate is indeed higher by almost a factor of $\sim$ 20 for the $H$ model compared to the value used for the standard model, which proves to be very destructive in the early stages. 
        
        This destruction comes from both the increased presence of ionised species and increased grain desorptions. H$^+$ and He$^+$ are two ions that play important roles in gaseous destruction. For example,  with CS, the main destruction route during the precollapse under the standard model comes from its photodissociation into S and C with some contribution from photoionization and reactions with ionised species. The increased ionization rates for the $L$ and the $H$ models result in increased abundances for the H$^+$ and the He$^+$ ions. The H$^+$ ion plays a part in the CS destruction through the route:
            
        \begin{equation*}
            {\rm H^+ + CS}\rightarrow{\rm CS^+ + H}
        \end{equation*}
          
        The importance of this reaction is enhanced with the $L$ model and is significantly more dominant for the $H$ model. The $H$ model also has higher destruction contributions from the He$^+$ ion via the routes:
        \begin{equation*}
            {\rm He^+ + CS}\rightarrow{\rm S^+ + C + He}
        \end{equation*}
        \begin{equation*}
            {\rm He^+ + CS}\rightarrow{\rm S + C^+ + He}
        \end{equation*}
            
        A second example of this is H$_2$O, where the destruction in the standard model has some contributions from C$^+$ and H$_2$O reactions but is mostly dominated by H$_2$O photodissociation into OH and H. Similarly to CS, the increased abundances of H$^+$  at higher ionization rates drove the destruction of H$_2$O through the reaction:
            
        \begin{equation*}
            {\rm H^+ + H_2O}\rightarrow{\rm H_2O^+ + H}
        \end{equation*}
        
        While gaseous species are seeing reduced abundances due to ions, solid species are seeing these reductions via CR induced desorptions. For example, H$_2$O\textsubscript{(grain)} in the standard model is primarily destroyed by CR induced UV desorptions. With the increased ionization rates of the $L$ model, direct CR desorption also begins to take place. The even higher ionization rates of the $H$ model result in both the CR induced UV desorption and the direct CR desorptions becoming more efficient, resulting in the reduced abundances seen.
        
        During the post-collapse, species only see notable abundance changes with the $H$ model. Gas phase H$_2$O and CS have increased abundances while the solid phase CO\textsubscript{(grain)} and CO$_2$\textsubscript{(grain)} show decreased abundances. In this case the solid phase species tend to undergo more destruction, with CO\textsubscript{(grain)} and CO$_2$\textsubscript{(grain)} decreasing in abundances by over 3 orders of magnitude and the gaseous species see increases up to a factor of 20. While these large decreases in abundances are mainly caused by CR induced desorptions, there are also contributions from reduced formation rates. CO freeze-out to CO\textsubscript{(grain)} is the dominant formation route during the cloud collapse. Under the $H$ model this formation method is significantly inhibited during the collapse phase, reducing the amount of freeze-out taking place. CO$_2$\textsubscript{(grain)} is also affected by the CO freeze-out inhibition. The primary formation route for CO$_2$\textsubscript{(grain)} comes from the diffusion of CO\textsubscript{(grain)} and OH\textsubscript{(grain)}. Reduced abundances for both of these species with the $H$ model inhibit the amount of CO$_2$\textsubscript{(grain)} formation. The combination of less formation and more desorption result in these significant decreases seen with the $H$ model. These increased desorptions can also influence the gas phase species. After the collapse, the primary H$^+$ destruction route for H$_2$O is no longer efficient; this fact coupled with the increased desorptions of H$_2$O\textsubscript{(grain)}, result in the increased postcollapse abundances seen. Other species, like CS, are not as reliant on desorptions. After the collapse the primary formation route for CS comes from the photodissociation of H$_2$CS into CS and H$_2$. The H$_2$CS molecule also shows increased abundances for the $L$ model and significantly so for the $H$ model, which in turn leads to more efficient photodissociation.
        
        When the final density is increased, this reduces the effects of the ionization dependency addition. CO\textsubscript{(grain)} and CO$_2$\textsubscript{(grain)} are prime examples of this effect, as the large decreases in abundances seen at 10$^4$ cm$^{-3}$ are no longer present at higher densities (abundance changes are now under an order of magnitude). The reduced ionization rates as density increases is the main cause of this feature through reduced CR desorptions. Also, under these conditions CO freeze-out during the collapse is not inhibited with the increased ionization rates. This, along with the lower desorptions, lead to the reduced effects of the ionization dependency.

        \subsubsection{Effects due to temperature variations}
        
            \begin{figure*}[tbp]
                \centering
                \includegraphics[width=18cm, height=12cm]{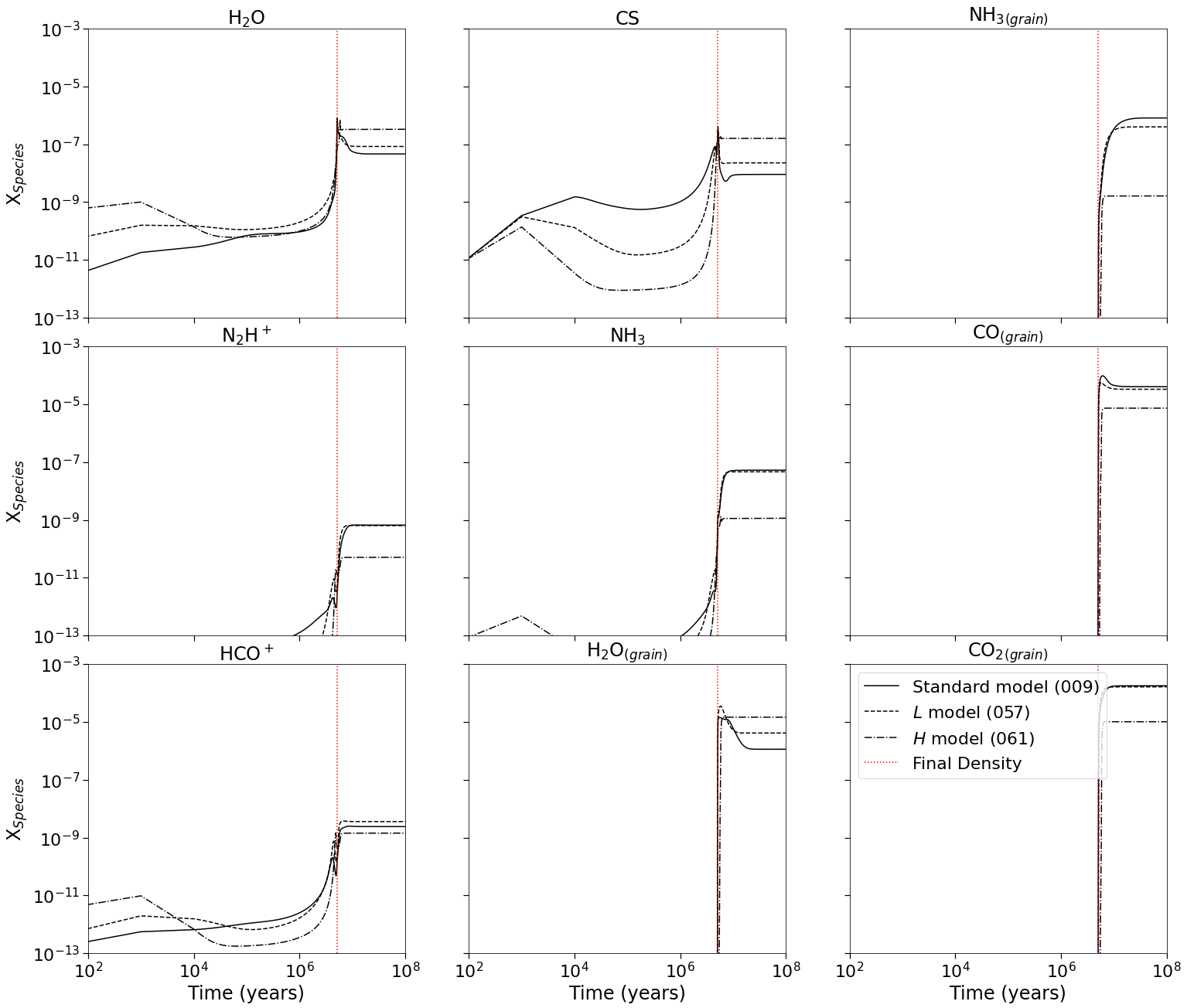}
                \caption{Plot showing the effect of $L$ and $H$ ionization models (short- and long-dashed black lines, respectively) on chemical abundances over time compared to the basic UCLCHEM handling (solid black line). These models have a final density of 10$^4$cm$^{-3}$, an initial temperature of 20 K, a radiation field of $\times$1 and a zeta of $\times$1.The red line represents the time which final density is reached and the numbers in the legend represent the model number identifier.}
                \label{fig:ion_1e4_T20}
            \end{figure*}

            Figure \ref{fig:ion_1e4_T20} shows the $L$ and $H$ models with an increased initial temperature of 20~K, while Table \ref{tab:ionBasLargeResults} summarises the effects on the $L$ and $H$ models with initial temperatures of 20~K  and 30~K. During the pre-collapse phase, when the temperature is increased, only CS and HCO$^+$ show notable changes. Both still show a reduction in abundance with the ionization dependency, but the reduction is lower at higher temperatures (CS is reduced by over 2 orders of magnitude and HCO$^+$ is reduced by up to a factor 6  for the $H$ model). At higher temperatures only H$_2$O, CS and HCO$^+$ have abundances above the set limit of 10$^{-13}$. For the post-collapse, increasing the temperature to 20 K results in larger abundance changes from the $H$ model for solid and gas phase NH$_3$. However, CS, CO\textsubscript{(grain)} and CO$_2$\textsubscript{(grain)} instead see less of a change than at 10~K (significant in the cases of CO\textsubscript{(grain)} and CO$_2$\textsubscript{(grain)}, where the large reduction in abundance at 10~K $H$ model is no longer seen). At 30~K only CS and NH$_3$ have their abundance change by over a factor of 3 (both by a factor of $\sim$8) and in both cases these changes are less than they are at 20~K for the $L$ model. Increasing the density at these temperatures has a similar effect as at 10~K (i.e. reduced changes as density increases).
            
            Gas phase NH$_3$ abundances are strongly influenced by the solid phase NH$_3$ abundances. Gas phase formation comes completely from grain desorptions. NH$_3$\textsubscript{(grain)} however, at 10~K under all models, is formed via H\textsubscript{(grain)} and NH$_2$\textsubscript{(grain)} diffusion. As temperature increases this reaction becomes less dominant, particularly during the pre-collapse phase and with the $H$ model at 30~K. In this case there is no formation from the diffusion reaction, which results in NH$_3$\textsubscript{(grain)} having abundances below the set limit.
            
           CO$_2$\textsubscript{(grain)}, at higher temperatures, relies less on CO\textsubscript{(grain)} diffusion. During the post-collapse the primary formation rate comes from the diffusion reaction:
            \begin{equation*}
                {\rm H_2CO\textsubscript{(grain)} + O\textsubscript{(grain)}}\rightarrow{\rm CO_2\textsubscript{(grain)} + H_2}
            \end{equation*}
            
            The reduced abundance changes for CO$_2$\textsubscript{(grain)} here are a result of the H$_2$CO\textsubscript{(grain)} abundances. Under the $L$ model, H$_2$CO\textsubscript{(grain)} has higher abundances, resulting in more diffusion, which is balanced out by the increased desorption of CO$_2$\textsubscript{(grain)}, leading to little change from the standard model. The $H$ model on the other hand, still sees a reduction in abundance, which is a result of the increased desorptions, but the reduction is not as severe as at 10~K due to the $H$ model having increased O\textsubscript{(grain)} abundances for more diffusion.

        \subsubsection{Effects due to variations in the radiation field strength}
        
            \begin{figure*}[tbp]
                \centering
                \includegraphics[width=18cm, height=12cm]{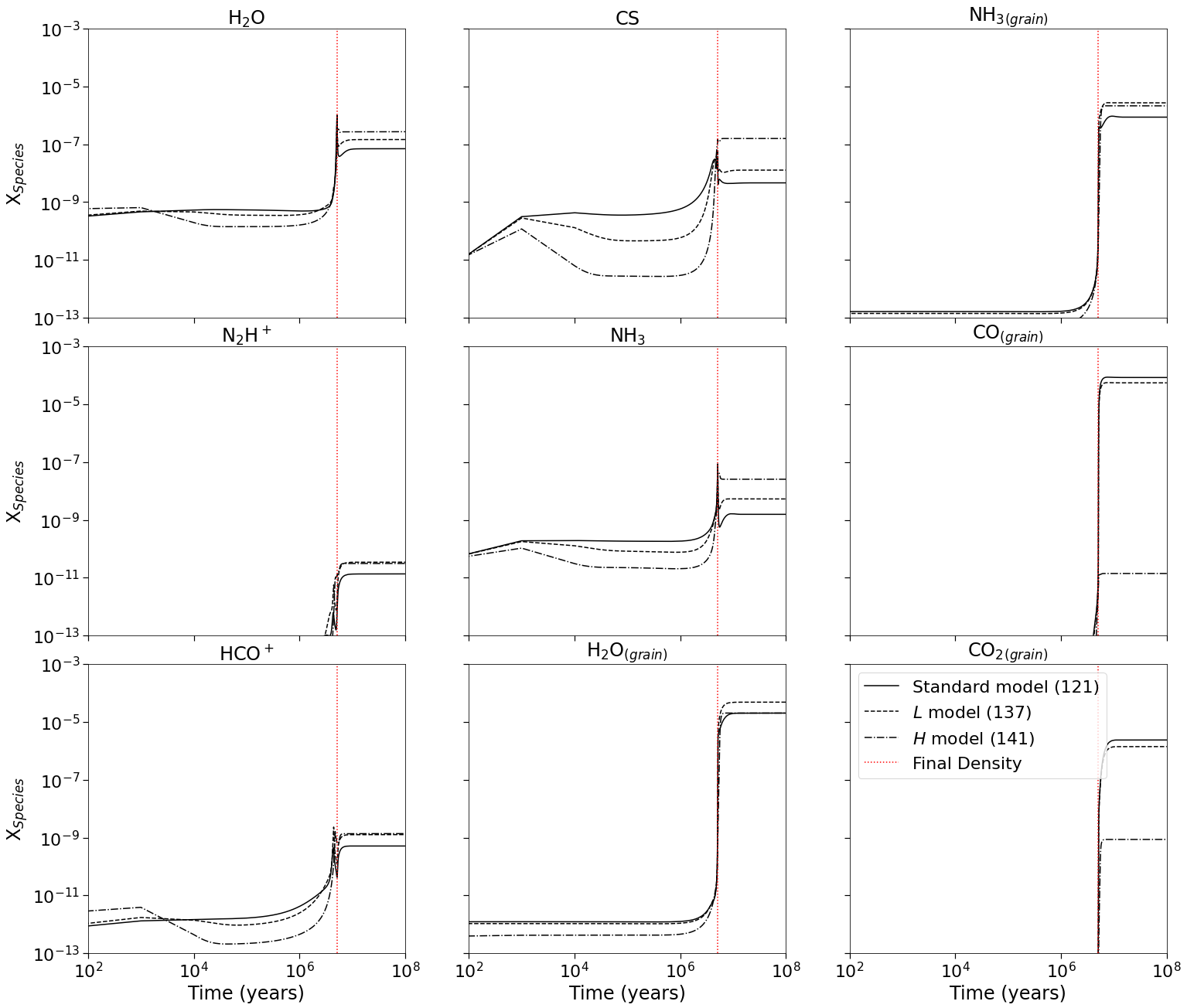}
                \caption{Plot showing the effect of $L$ and $H$ ionization models (short- and long-dashed black lines, respectively) on chemical abundances over time compared to the basic UCLCHEM handling (solid black line). These models have a final density of 10$^4$ cm$^{-3}$, an initial temperature of 10K, a radiation field of $\times$10 and a zeta of $\times$1. The red line represents the time which final density is reached and the numbers in the legend represent the model number identifier.}
                \label{fig:ion_1e4_R10}
            \end{figure*}
        
            Figure \ref{fig:ion_1e4_R10} shows the differences between the standard model and the models where the new treatment of the cosmic ionization rate is included, when the radiation field is increased by a factor of 10. Table \ref{tab:ionBasLargeResults} summarizes the results for enhancing the radiation field by a factor of 10 and 100.
            
            In the pre-collapse phase, enhancing the radiation factor by a factor of 10 reduces the abundance changes produced by the ionization dependency, which are further reduced when the radiation field is increased by a factor of 100 (abundance changes are up to 1 order of magnitude, see Table \ref{tab:ionBasLargeResults}). During the post-collapse phase, in general, increasing the radiation field enhances the effects of the ionization rate dependency. At a radiation field of 100 Habing only H$_2$O, CS, NH$_3$, and HCO$^+$ are above the $10^{-13}$ threshold and have increased abundances compared to the standard model. H$_2$O, NH$_3$ and HCO$^+$ changes are enhanced by $\sim$1, $\sim$2 and $\sim$1 orders of magnitude, respectively. The large abundance increase for NH$_3$ comes from desorption from the grains. Under the standard and the $L$ model, NH$_3$\textsubscript{(grain)} abundances are below the set limit. This is not the case for the $H$ model. The increased grain abundance here is due to the diffusion of H\textsubscript{(grain)} and NH$_2$\textsubscript{(grain)}. Under these conditions, both of these species have significantly higher abundances with the $H$ model than the $L$ or standard. This increases NH$_3$\textsubscript{(grain)} formation which then can desorb into the gas phase. For HCO$^+$, under the increased radiation field there are two main formation routes, photoionization of HCO and the H$_3^+$ reaction:
            \begin{equation*}
                {\rm H_3^+ + CO }\rightarrow{\rm HCO^+ + e^-}
            \end{equation*}
            
            Under the $L$ and $H$ models, both CO and H$_3^+$ have much higher abundances, leading to the increased production of HCO$^+$. When the density is increased above 10$^4$ cm$^{-3}$, there are no significant differences between the effects of the ionization dependency at the standard radiation field strength and the effects at increased strengths.
            
        \subsubsection{Effects due to the initial CR Ionization rate variations}
        
            \begin{figure*}[tbp]
                \centering
                \includegraphics[width=18cm, height=12cm]{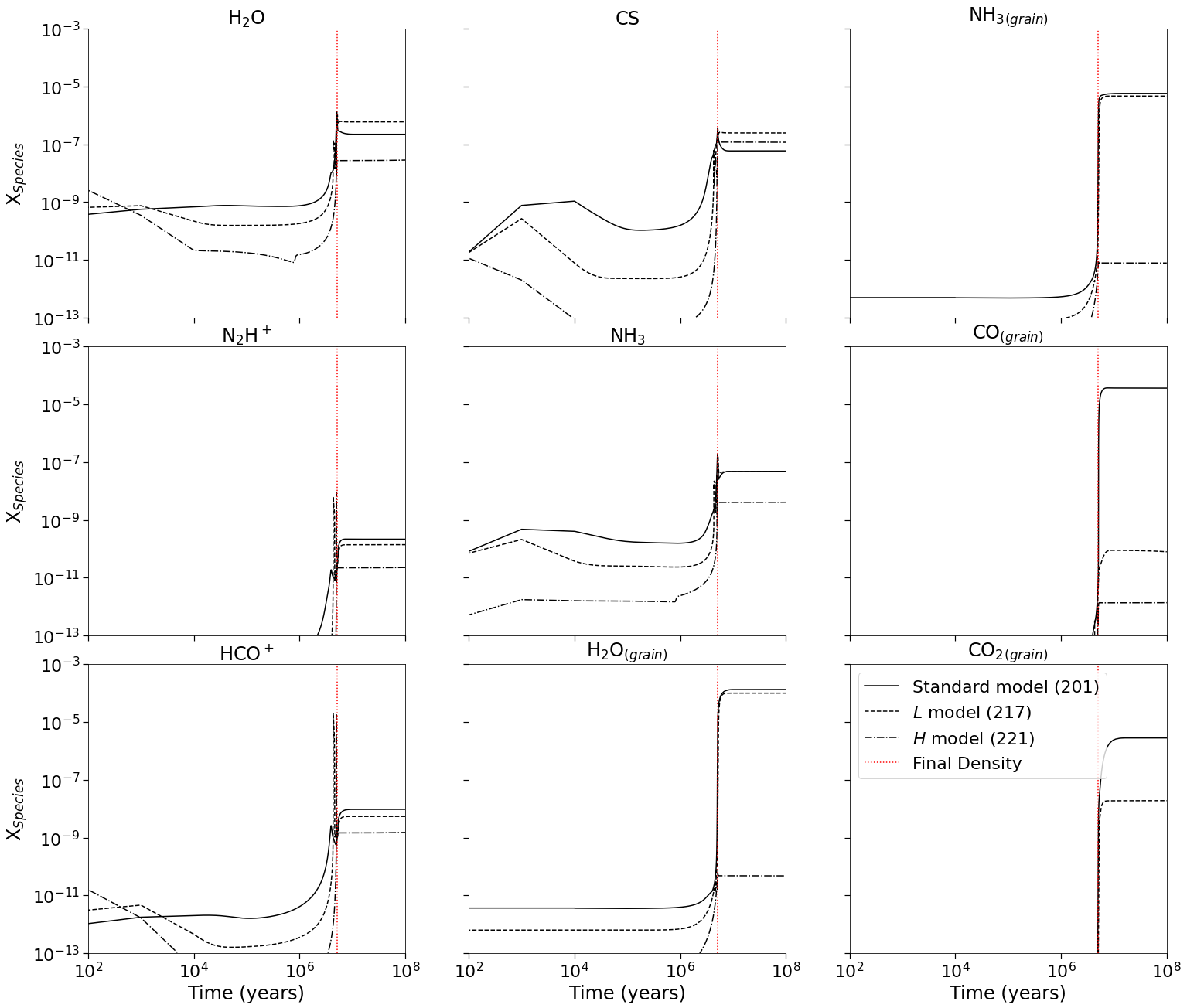}
                \caption{Plot showing the effect of $L$ and $H$ ionization models (short- and long-dashed black lines, respectively) on chemical abundances over time compared to the basic UCLCHEM handling (solid black line). These models have a final density of 10$^4$ cm$^{-3}$, an initial temperature of 10K, a radiation field of $\times$1 and a zeta of $\times$10. The red line represents the time which final density is reached and the numbers in the legend represent the model number identifier.}
                \label{fig:ion_1e4_Z10}
            \end{figure*}
        
            Figure \ref{fig:ion_1e4_Z10} shows the effects of increasing the ionization rate by a factor of 10 on the chemical abundances with the ionization rate dependency. Table \ref{tab:ionBasLargeResults} summarizes the results of increasing the rate by factors of 10 and 100.
            
            Increasing the initial ionization rate in this manner proves to be very destructive both with and without the ionization dependency, particularly in the pre-collapse period. Many species have abundances below the set limit with this increased ionization rate factors. Those that are visible show much larger abundance reductions (up to 2 orders of magnitude, see Table \ref{tab:ionBasLargeResults}) than at an un-adjusted rate. During the post-collapse phase, increasing the ionization rate by a factor of 10 and 100, increases the influence of the ionization dependency. Species that had no notable changes under the standard rate now show reduced abundances. CO\textsubscript{(grain)} and CO$_2$\textsubscript{(grain)} again are particularly affected, showing reductions of up to 7 orders of magnitude. These effects are more enhanced with the $\times$100 initial ionization rate. As with the other conditions, cores of higher densities show reduced effects from the ionization dependency, even with the enhanced initial CR ionization rate. These trends originate for the same reasons as for the models with a standard initial CR ionization rate but are more pronounced (e.g. increased ions and desorption).

    \subsection{Density Dependent Dissociation Rates}

        Table \ref{tab:dissexcitedResults} summarises the only cases where the H$_2$ dissociation rate has any effect. In short, under an enhanced radiation field of 100 times the galactic one, an increase of abundance was seen for the solid species (by a factor of $\sim$4), at a low density (10$^4$ cm$^{-3}$) and only for the $L$ model. The other notable effect is seen at an increased initial ionization rate of $\times$10, for CO\textsubscript{(grain)}, also at 10$^4$ cm$^{-3}$ with the $L$ model, where a significant increase of abundance ($\sim$3 orders of magnitude) is seen. This large increase in abundance can be traced to the diffusion of H\textsubscript{(grain)} and CO\textsubscript{(grain)} into HCO\textsubscript{(grain)}. Under these conditions this is the dominant destruction route for CO\textsubscript{(grain)}. Under the H$_2$ dissociation rate this reaction pathway is severely inhibited, reducing the destruction of CO\textsubscript{(grain)} during and after the collapse. Abundances of H\textsubscript{(grain)} here are also lower for the H$_2$ dissociation dependency model, which may explain the inhibition.

    \subsection{Excited Species}
        Table \ref{tab:dissexcitedResults} summarises the only conditions where the inclusion of the excited species had a notable effect on the abundances of the selected species. At an increased temperature of 30~K, N$_2$H$^+$ shows a reduced abundance by a factor of 4 at a density of 10$^5$ cm$^{-3}$. Increasing the initial ionization rate by a factor of 10 reduces CO$_2$\textsubscript{(grain)} abundances at a density of 10$^4$ cm$^{-3}$. The most significant effects come from increasing the ionization by a factor of 100. While the higher ionization rates provide more excitations, the increased destruction of the species is not only from their excitation and subsequent reactions. CO\textsubscript{(grain)} for example is also affected heavily by the H\textsubscript{(grain)} and CO\textsubscript{(grain)} diffusion reaction. The addition of the excited species also produce higher abundances of H\textsubscript{(grain)}, which increased the amount of CO\textsubscript{(grain)} diffusion. 
        
    \subsection{Density Dependency with Excited Species}
    
        In this section models with both the ionization rate and dissociation rate dependencies activated are compared with and without the inclusion of excited species. Table \ref{tab:combinedResults} show the effects under "standard" conditions (10 K initial temperature, radiation field strength of 1 Habing and an initial ionization factor of $\times$1). CO\textsubscript{(grain)} and CO$_2$\textsubscript{(grain)} are the main species affected, and have reduced abundances when the excited species are included. These reduced abundances are caused via the same destructive methods as discussed in the previous subsection.

        \subsubsection{Effects due to parameter variations}
        
            When the excited species are included with the CR ionization and H$_2$ dissociation in the chemical models, the effects of varying the temperature and radiation field strength are reduced, while the effects of varying the ionization factor are increased. As such these effects are quickly summarised here.
            
            Increasing the initial temperature and increasing the radiation field strength both inhibit the effects of including the excited species. Under a higher temperature, including the excited species only has an effect on N$_2$H$^+$ (this is at 30~K and a density of 10$^7$ cm$^{-3}$ with the $H$ model only) where the species shows an increased abundance. Increasing the radiation field strength only has an inhibition effect on lower densities. At 10$^5$ cm$^{-3}$ and above, there are no differences between 1, 10 and 100 Habing.
            
            Including the excited species with increased initial ionization rates of 10 and 100 times standard, has a greater effect than at the $\times$1 value. The abundance changes are both larger and seen for more species. At a 10 times standard initial ionization rate, effects are only seen with the $H$ model at 10$^5$ cm$^{-3}$ and 10$^6$ cm$^{-3}$. In these conditions, most species see reduced abundances (several orders of magnitude for CS, CO\textsubscript{(grain)}, HCO$^+$ and CO$_2$\textsubscript{(grain)}). Under an increased rate of 100 times standard, the species see larger reductions, under the same densities but with the $L$ model instead. Similar reductions are also seen here at 10$^7$ cm$^{-3}$ with both the $L$ and $H$ models.

        \begin{table*}[tbp]
            \centering
            \begin{tabular}{|m{2.5cm}| m{5cm}| m{9cm}|}
                \hline
                \textbf{Species} & \textbf{Conditions} & \textbf{Behaviour} \\
                \hline
                \hline
                \multicolumn{3}{|c|}{\textit{Ionization rate dependency}} \\
                \hline
                H$_2$O & 10$^6$ cm$^{-3}$, $H$ model  & Abundance increase larger than other densities. \\
                \hline
                N$_2$H$^+$ & 10$^5$ cm$^{-3}$, $H$ model & Only density to see a notable change (increased abundance). \\
                \hline
                NH$_3$ & 10$^6$ cm$^{-3}$, $H$ model & Only density to see a notable change (increased abundance). \\
                \hline
                HCO$^+$ & 10$^6$ cm$^{-3}$, $H$ model & Abundance increase larger than other densities. \\
                \hline
                \hline
                \multicolumn{3}{|c|}{\textit{Ionization rate dependency with parameter variations}} \\
                \hline
                N$_2$H$^+$ & 20~K,  10$^6$ cm$^{-3}$, $H$ model & Abundance change greater than other densities.\\
                & 30~K, 10$^6$ cm$^{-3}$, $H$ model & Only condition to see abundance change at 30 K. Abundance not as reduced as at 20~K. \\
                \hline
                NH$_3$\textsubscript{(grain)} & 30~K, 10$^5$ cm$^{-3}$, $H$ model & Only condition to see abundance change at 30~K. Abundance further reduced than at 20~K. \\
                \hline
                CO$_2$\textsubscript{(grain)} & 30~K, 10$^5$ cm$^{-3}$, $H$ model & Only condition to see abundance change at 30~K. Abundance further reduced than at 20~K. \\
                \hline
                CS & 100 times radiation field, 10$^4$ cm$^{-3}$, $H$ model & Abundance is not as reduced as at 10 times radiation field. \\
                \hline
                H$_2$O\textsubscript{(grain)} & 100 times initial ionization rate, 10$^5$ cm$^{-3}$, $H$ model & Reduction in abundance is greater than at 10$^4$ cm$^{-3}$. \\
                \hline

                \end{tabular}
            \caption{Table showing the post-collapse species, conditions and the behaviour that do not follow the general trends of the ionization dependency.}
            \label{table: outlier table}
        \end{table*}   

    \subsection{Comparison to observations}
 
        In this section, we qualitatively compare our models to a set of observations from the Cyanopolyyne peak ("CP" or Core D; \cite{Hirahara_1992}) of the molecular cloud TMC-1 which is thought to currently undergoing rapid core formation \citep{Choi2017}. This core was chosen as it has been well studied and its density is expected to be around 10$^4$ cm$^{-3}$ with a temperature of about 10~K. The density of 10$^4$ cm$^{-3}$ is a good candidate for this study, as the effects of the ionization rate dependency are greater due to the low density.
        
        \begin{figure*}[tbp]
            \centering
            \includegraphics[width=0.49\textwidth]{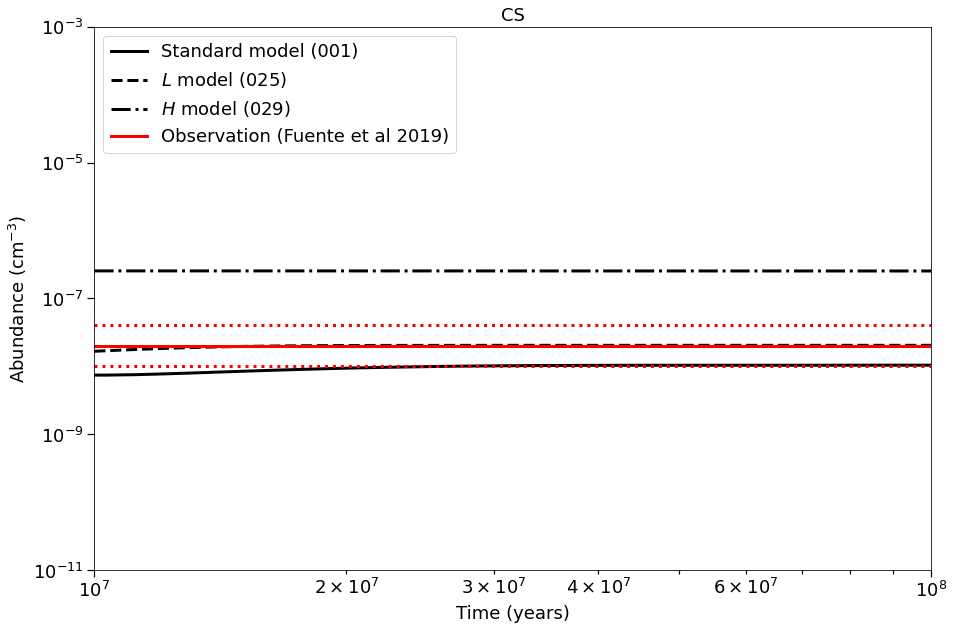}
            \includegraphics[width=0.49\textwidth]{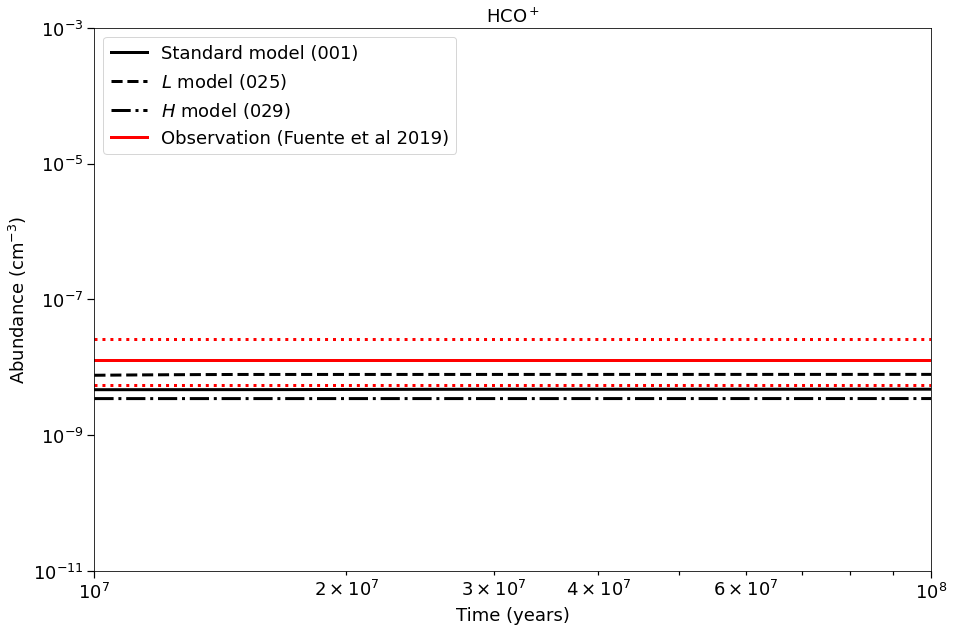} \\
            \includegraphics[width=0.49\textwidth]{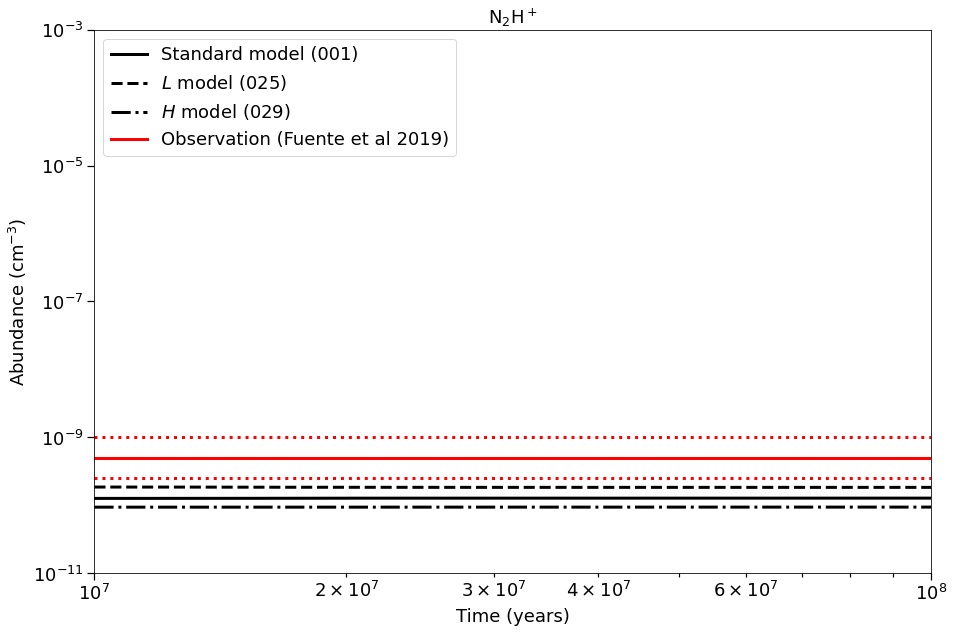}

            \caption{Figure showing the upper and lower limits to species in TMC-1, derived from \cite{Fuente2019}. Figures compare observed abundances of CS (top left), HCO$^+$ (top right) and N$_2$H$^+$ (bottom) limits to UCLCHEM post-collapse models. The red dotted lines show the upper and lower limits.}
    
            \label{fig:observation_comparison_fuente}
        \end{figure*}
 
        \begin{figure*}[h]
            \centering
            \includegraphics[width=0.49\textwidth]{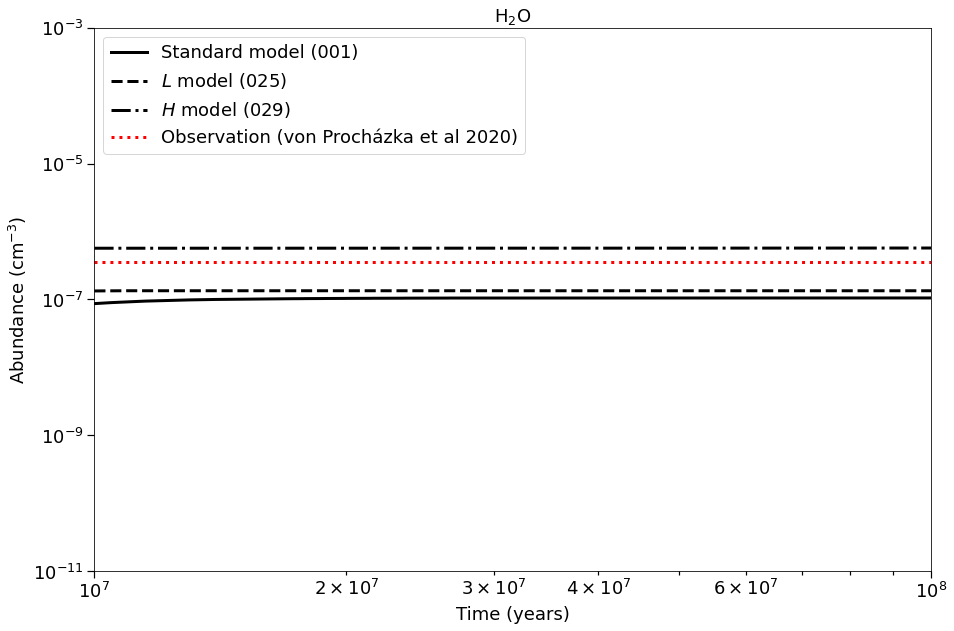}
            \includegraphics[width=0.49\textwidth]{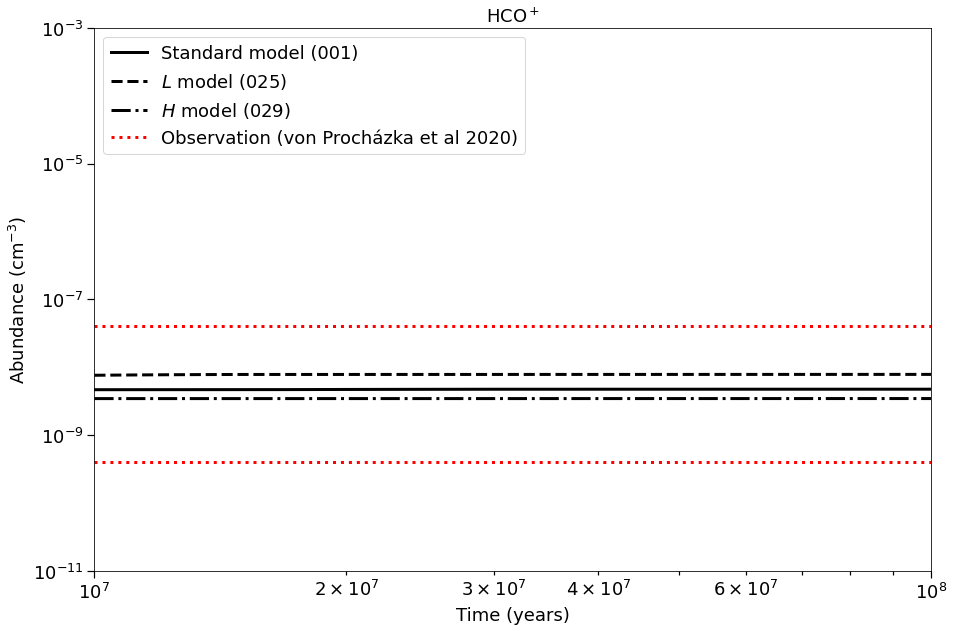} \\
            \includegraphics[width=0.49\textwidth]{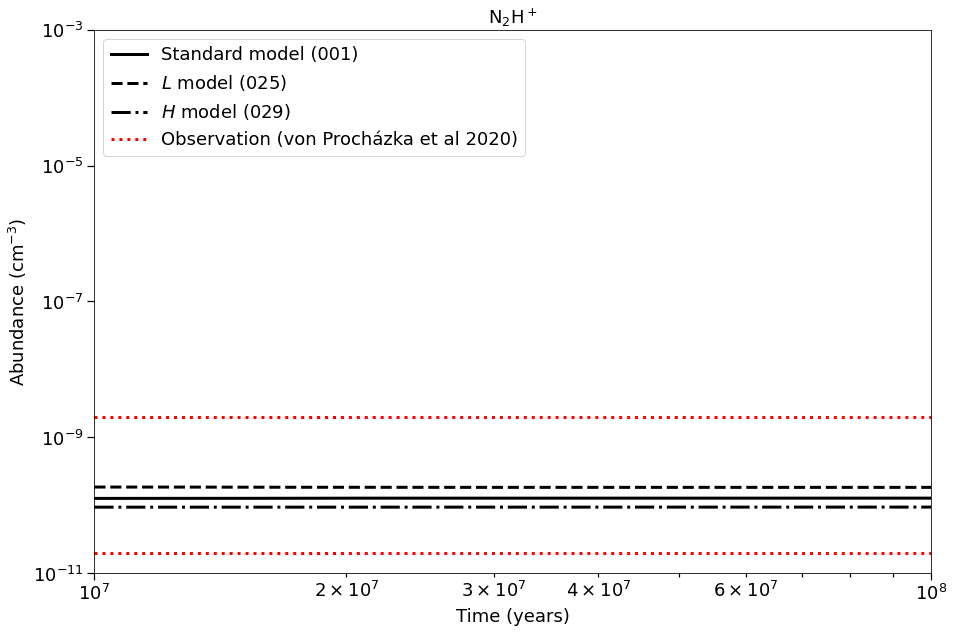}
            \includegraphics[width=0.49\textwidth]{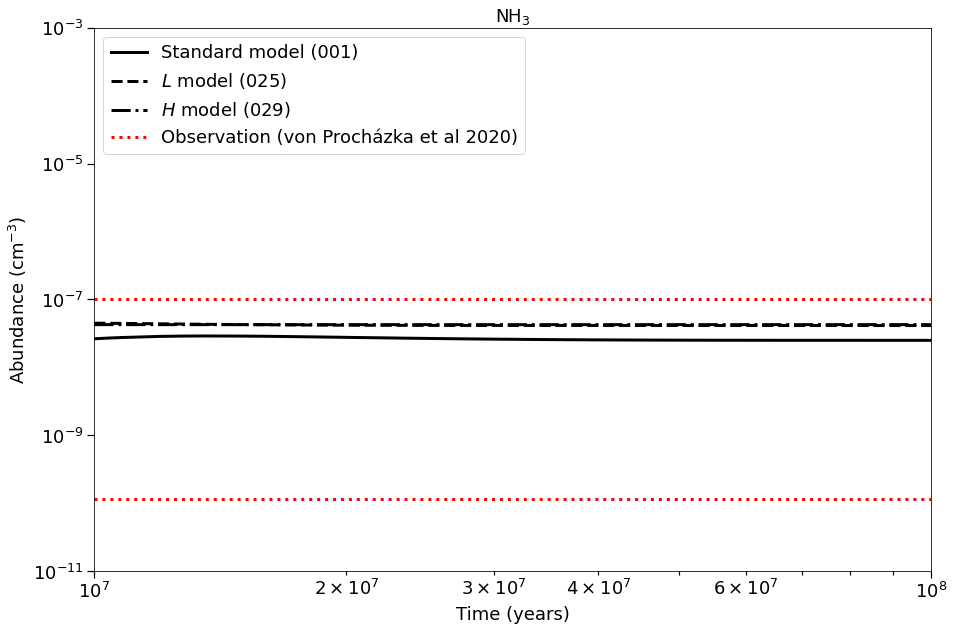}       
    
            \caption{Figure showing the upper and lower limits to species in TMC-1, derived from \cite{von_Proch_zka_2020}. Figures compare observed abundances of H$_2$O (top left), HCO$^+$ (top right), N$_2$H$^+$ (bottom left) and NH$_3$ (bottom right) limits to UCLCHEM post-collapse models. The red dotted lines show the upper and lower limits (for H$_2$O there are only upper limits).}
    
            \label{fig:observation_comparison}
        \end{figure*}
        
        Both \textcite{von_Proch_zka_2020} and \textcite{Fuente2019} report chemical abundances of the CP region. \textcite{Fuente2019} reports molecular abundances derived from observations using the IRAM 30m (3mm and 2mm) and the Yebes 40m telescopes. Depending on the setup, the IRAM 30m has a spatial resolution of $\sim$29" and the Yebes 40m has a HPBW of 42" or 84". On the other hand, \textcite{von_Proch_zka_2020} reports a collection of upper and lower abundance limits obtained from $\sim$20 other studies (see Table 2 in \textcite{von_Proch_zka_2020}), with the emphasis that focusing on upper and lower limits somewhat mitigates the errors in observations and modelling.
        
        Figure \ref{fig:observation_comparison_fuente} shows the data reported in \cite{Fuente2019}, compared to the UCLCHEM models at 10 K and a final density of 10$^4$ cm$^{-3}$ with and without the inclusion of the CR ionization rate dependency during the post-collapse phase. The "standard" UCLCHEM model for CS and HCO$^+$ tends to display abundances that are near the lower limit, while the $L$ is closer to the central values. The $H$ model over predicts the abundance for CS and under predicts the abundance for HCO$^+$. With N$_2$H$^+$ all models under predict the abundance, with the $L$ model being the closest and the $H$ model under predicting the most.\\
    
        Figure \ref{fig:observation_comparison} shows data reported in \textcite{von_Proch_zka_2020}, also compared to the UCLCHEM models at 10 K and a final density of 10$^4$ cm$^{-3}$ with and without the inclusion of the CR ionization rate dependency during the post-collapse phase. In the cases of HCO$^+$, N$_2$H$^+$ and NH$_3$, for all models the post-collapse abundances are within in the upper and lower limits of the observations. In \textcite{von_Proch_zka_2020} only an upper limit is noted for H$_2$O. In this case the $H$ model exceeds the stated limit while the $L$ and basic model do not. \\
        
        In the case of this region in TMC-1 and the compared species, the $L$ model of the ionization dependency appears to perform the best of the three models. The $H$ model over or under predicts the abundances on several occasions, suggesting this upper limit for the ionization rate may in fact be too high. \\
        
        One caveat that must be addressed is how UCLCHEM handles grains. This version of UCLCHEM considers a grain to be a single layer (i.e. no distinction between grain surface and bulk). It is therefore necessary to speculate on the effects a multi-layer grain approach may have on these results. Species in the bulk are somewhat shielded from CR impacts and the subsequent desorptions. As CR desorptions are critical to abundance changes seen for species like CO$_{(grain)}$ and CO$_{2(grain)}$, it is likely the abundance changes seen with the inclusion of the CR ionization dependency will be less significant. Excited species in the bulk are also more protected. Desorptions from excitations and excited reactions would be reduced with a greater emphasis on relaxations, again reducing the effects we see in our models.
\section{Summary}
\label{summary}

    In this paper we improve the treatment of cosmic rays in the gas-grain time dependent chemical code UCLCHEM by including the dependency of the cosmic ray ionization and H$_2$ dissociation rates on the column density of the gas, as well as the excited species due to the cosmic rays on the grains. We then evaluate the effects of these additions on the  chemistry of pre-stellar cores. It is evident that the cosmic ray ionization rate dependency on the column density of the core is the most influential of the treatments, with the inclusion of excited species on the grains playing roles only under specific conditions. Our conclusions can be summarized as follows:
    
    \begin{itemize}
        \item In the low densities of the precollapse phase ($\sim$10$^2$ cm$^{-3}$) the ionization rate dependency is very destructive due to CR induced desorptions and the production of chemically important ions.
        
        \item After the core collapses, the inclusion of the dependency of the CR ionization rate on the column density of the core leads to increased grain desorptions, which decreases solid species abundances (and subsequently increases gaseous species abundances), species like H$_2$O, CO\textsubscript{(grain)} and CO$_2$\textsubscript{(grain)} are particularly affected by this. Other gaseous species, like CS show increased abundances from dissociations of larger molecules like H$_2$CS.
        
        \item Changing the physical parameters of the cloud alters the impact of the new treatments in a non-trivial manner. Higher densities have lower ionization rates with the dependency, reducing the abundance changes seen for all species. Increasing the temperature also has a similar effect for CS, CO\textsubscript{(grain)} and CO$_2$\textsubscript{(grain)} (increased formation rates balance out the destruction from the CRs), while NH$_3$ shows lower abundances due to less NH$_3$\textsubscript{(grain)} formation and subsequent desorption. Increasing the radiation field strength enhances the effects of the ionization dependency, which occurs as a result of grain and gas formation routes. NH$_3$ for example sees increased abundances with the $H$ model due to higher NH$_3$\textsubscript{(grain)} grain formation and desorption, while HCO$^+$ sees larger changes due to formation in the gas via H$_3^+$ and CO.

        \item The H$_2$ dissociation rate dependency and the inclusion of excited species only affect the chemistry of some of the investigated species under specific conditions. The H$_2$ dissociation dependency increases abundances of some solid species for the $L$ model under two conditions, a $\times$10 ionization rate and a $\times$100 radiation field strength. CO\textsubscript{(grain)} sees these abundance increase due to the inhibition of its primary destruction route, from a reduced abundance of the H\textsubscript{(grain)} reactant. The excited species, reduce solid abundances at higher ionization rates, particularly with CO$_2$\textsubscript{(grain)} and CO\textsubscript{(grain)}. While the excitations and subsequent reactions reduced the solid abundances, destruction also comes from reactions with H\textsubscript{(grain)} (which also sees higher abundances under these conditions).
        
        \item Chemical models with and without the ionization dependency were compared to molecular abundances in the TMC-1 cyanopolyyne peak from \textcite{Fuente2019} and \textcite{von_Proch_zka_2020}. The comparisons show that the $L$ model of the dependency tends to reproduce abundances more reliably than the standard handling or the $H$ model. All models had more difficulty reproducing N$_2$H$^+$ abundances (the $L$ model still performed better). The $H$ model predicted abundances outside the \textcite{Fuente2019} observational limits for both CS and HCO$^+$ and over predicted the abundance of H$_2$O compared to \textcite{von_Proch_zka_2020}.
    \end{itemize}

\section{Acknowledgements}
We would like to thank Jonathan Holdship for his contribution to the chemical modelling, particularly with UCLCHEM's analysis tools. This project has received funding from the European Union’s Horizon 2020 research and innovation programme under the Marie Skłodowska-Curie grant agreement No 811312 for the project "Astro-Chemical Origins” (ACO). This work is part of a project that has received funding from the European Research Council (ERC) under the European Union’s Horizon 2020 research and innovation programme MOPPEX 833460.

\newpage

\bibliography{bibliography.bib}{}

\bibliographystyle{aasjournal}

\newpage

\appendix

\setcounter{table}{0}
\renewcommand{\thetable}{A\arabic{table}}
    
    \begin{table}[!htbp]
        \centering
        \begin{tabular}{|c|c|c|}
            \hline
            \multicolumn{3}{|c|}{CR ionization dependency coefficients}\\
            \hline
            $k$ & $c_k$ (model $L$) & $c_k$ (model $H$)  \\
            \hline
            0 &	$1.545456645800 \times 10 ^{7}$ & $1.223529865309 \times 10 ^{7}$\\
            \hline
            1 &	$-6.307708626617 \times 10 ^{6}$ & $-5.013766644305 \times 10 ^{6}$\\
            \hline
            2 &	$1.142680666041 \times 10 ^{6}$ &	$9.120125566763 \times 10 ^{5}$ \\
            \hline
            3 &	$-1.205932302621 \times 10 ^{5}$ & $-9.665446168847 \times 10 ^{4}$ \\
            \hline
            4 &	$8.170913352693 \times 10 ^{3}$ &	$6.576930812109 \times 10 ^{3}$ \\
            \hline
            5 &	$-3.686121296079 \times 10 ^{2}$ & $-2.979875686226 \times 10 ^{2}$ \\
            \hline
            6 &	$1.107203722057 \times 10 ^{1}$ &	$8.989721355058 \times 10 ^{0}$ \\
            \hline
            7 &	$-2.135293914267 \times 10 ^{-1}$ & $-1.741300519598 \times 10 ^{-1}$ \\
            \hline
            8 &	$2.399219033781 \times 10 ^{-3}$ & $1.965098116126 \times 10 ^{-3}$ \\
            \hline
            9 &	$-1.196664901916 \times 10 ^{-5}$ & $-9.844203439473 \times 10 ^{-6}$ \\
            \hline
            \hline
            \multicolumn{3}{|c|}{CR H$_2$ dissociation dependency coefficients}\\
            \hline
            $k$ & $c_k$ (model $L$) & $c_k$ (model $H$)  \\
            \hline
            0 &	$1.582911005330 \times 10 ^{7}$ & $1.217227462831 \times 10 ^{7}$\\
            \hline
            1 &	$-6.465722684896 \times 10 ^{6}$ & $-4.989649250304 \times 10 ^{6}$\\
            \hline
            2 &	$1.172189025424 \times 10 ^{6}$ & $9.079152156645 \times 10 ^{5}$ \\
            \hline
            3 &	$-1.237950798073 \times 10 ^{5}$ & $-9.624890825395 \times 10 ^{4}$ \\
            \hline
            4 &	$8.393404654312 \times 10 ^{3}$ &	$6.551161486120 \times 10 ^{3}$ \\
            \hline
            5 &	$-3.788811358130 \times 10 ^{2}$ &	$-2.968976216187 \times 10 ^{2}$ \\
            \hline
            6 &	$1.138688455029 \times 10 ^{1}$ &	$8.959037875226 \times 10 ^{0}$ \\
            \hline
            7 &	$-2.197136304567 \times 10 ^{-1}$ &	$-1.735757324445 \times 10 ^{-1}$\\
            \hline
            8 &	$2.469841278950 \times 10 ^{-3}$ &	$1.959267277734 \times 10 ^{-3}$ \\
            \hline
            9 &	$-1.232393620924 \times 10 ^{-5}$ &	$-9.816996707980 \times 10 ^{-6}$ \\
            \hline
            
        \end{tabular}
        \caption{Coefficients from \cite{Padovani_2018-ion} and \cite{Padovani_2018-diss} for the rate dependencies}
        \label{tab:coefficients}
    \end{table}

    \begin{table}[!htbp]  
    \fontsize{10pt}{10pt}
    \centering{\small

        \begin{tabular}{m{1.1cm} m{1.4cm}m{1.4cm}m{0.2cm}m{0.6cm}m{1.4cm}m{0.6cm}m{1.4cm}m{0.6cm}m{1.4cm}m{0.7cm}m{0.7cm}}
            \hline \hline
            &  \multicolumn{2}{c}{Pre-Collapse} & & \multicolumn{7}{c}{Post-Collapse} \\
            \hline
            Density & \multicolumn{2}{c}{10$^2$ cm$^{-3}$} & & \multicolumn{2}{c}{10$^4$ cm$^{-3}$} & \multicolumn{2}{c}{10$^5$ cm$^{-3}$} & \multicolumn{2}{c}{10$^6$ cm$^{-3}$} & \multicolumn{2}{c}{10$^7$ cm$^{-3}$}\\
            \hline
            Species &  L & H & & L & H & L & H & L & H & L & H\\
            \hline
                
            H$_2$O & --- & $-9.5$ & & --- & $+5.4$ & --- & $+3.6$ & --- & $+6.1$ & --- & --- \\
                
            CS & $-6.2\times10^1$ & $-1.1\times10^3$ & &  --- & $+2.4\times10^1$ & --- & $+1\times10^1$ & --- & $+1.2\times10^1$ & --- & $+7.1$ \\
                
            NH$_3$\textsubscript{(grain)} & --- & $-7$ & & --- & --- & --- & --- & --- & --- & --- & --- \\
                
            N$_2$H$^+$ & $<$ limit & $<$ limit & & --- & --- & --- & $+6.8$ & --- & --- & --- & --- \\
                
            NH$_3$ & $-6.8$ & $-2.9\times 10^1$ & & --- & --- & --- & --- & --- & $+5.7$ & --- & ---\\
                
            CO\textsubscript{(grain)} & $<$ limit & $<$ limit && --- & $-4.4\times 10^6$ & --- & --- & --- & $+3.5$ & --- & --- \\
                
            HCO$^+$ & $-4.2$ & $-2.4\times 10^1$ && --- & --- & --- & $+4.8$ & --- & $+8.2$ & --- & --- \\
                
            H$_2$O\textsubscript{(grain)} & --- & $-6.8$ && --- & --- & --- & --- & --- & --- & --- & --- \\
                
            CO$_2$\textsubscript{(grain)} & $<$ limit & $<$ limit && --- & $-5.3\times 10^3$ & --- & $-6.1$ & --- & --- & --- & $-5.7$ \\

            \hline
            \hline

        \end{tabular}}
        \caption{Table showing results of the ionization rate dependency on abundances. Values shown are representing increase($+$) or decrease($-$) of abundances compared to the basic model and by what factor the abundances differed. This table only shows abundance changes greater than a factor of 3 and any value marked "$<$ limit" is below the lower abundance limit of 10$^{-13}$.}
        \label{tab:ionBasResults}
    \end{table}

    \begin{sidewaystable*}[!htbp]
        \centering
        \tiny

        \begin{tabular}{m{1.5cm} m{1.5cm} m{1.4cm}m{1.4cm}m{0.2cm}m{1.2cm}m{1.4cm}m{1.2cm}m{1.4cm}m{1.2cm}m{1.4cm}m{1.2cm}m{1.4cm}}
            \hline \hline
            & &  \multicolumn{2}{c}{Pre-Collapse} & & \multicolumn{7}{c}{Post-Collapse} \\
            \hline
            & Density: & \multicolumn{2}{c}{10$^2$ cm$^{-3}$} & & \multicolumn{2}{c}{10$^4$ cm$^{-3}$} & \multicolumn{2}{c}{10$^5$ cm$^{-3}$} & \multicolumn{2}{c}{10$^6$ cm$^{-3}$} & \multicolumn{2}{c}{10$^7$ cm$^{-3}$}\\
            Temperature (K) & Species & L & H & & L & H & L & H & L & H & L & H\\
            \hline
            20 & H$_2$O & --- & --- & & --- & $+7.2$ & --- & $+6.8$ & --- & $+5.5$ & --- & --- \\

            20 & CS & $-3.8\times 10^1$ & $-6.7\times 10^2$ & & --- & $+1.8\times 10^1$ & --- & $+8.2$ & --- & $+8.7$ & --- & $+4.3$ \\

            20 & NH$_3$\textsubscript{(grain)} & $<$ limit & $<$ limit & & --- & $-5\times 10^2$ & --- & $-3.1$ & --- & --- & --- & --- \\

            20 & N$_2$H$^+$ & $<$ limit & $<$ limit & & --- & $-1.3\times 10^1$ & --- & $+5.2$ & --- & $+1.9\times 10^1$ & --- & $+3.8$ \\

            20 & NH$_3$ & $<$ limit & $<$ limit & & --- & $-4.6\times 10^1$ & --- & $+7.6$ & --- & $+3.3$ & --- & --- \\

            20 & CO\textsubscript{(grain)} & $<$ limit & $<$ limit & & --- & $-5.5$ & --- & $+5.9$ & --- & $+6$ & --- & $+3.8$ \\

            20 & HCO$^+$ & --- & $-5.9$ & & --- & --- & --- & --- & --- & --- & --- & --- \\

            20 & H$_2$O\textsubscript{(grain)} & $<$ limit & $<$ limit & & $+3.6$ & $+1.3\times 10^1$ &--- & --- & --- & --- & --- & --- \\
            
            20 & CO$_2$\textsubscript{(grain)} & $<$ limit & $<$ limit & &--- & $-1.8\times 10^1$ & --- & --- & --- & --- & --- & --- \\
                   
            30 & H$_2$O & --- & --- & & --- & --- & --- & --- & --- & $+8.4$ & --- & --- \\
            
            30 & CS & $-3.6\times 10^1$ & $-5.2\times 10^2$ & & --- & $+7.8$ & --- & $+6$ & --- & $+5.3$ & --- & $+4$ \\
            
            30 & NH$_3$\textsubscript{(grain)} & $<$ limit & $<$ limit & & --- & ---& --- & $-1.5\times 10^2$ & --- & --- & --- & --- \\
            
            30 & N$_2$H$^+$ & $<$ limit & $<$ limit & & --- & --- & --- & --- & --- & $+1\times 10^1$ & --- & --- \\
            
            30 & NH$_3$ & $<$ limit & $<$ limit & & --- & $-8.1$ & --- & --- & --- & $+7$ & --- & --- \\     
            
            30 & HCO$^+$ & --- & $-4.4$ & & --- & --- & --- & --- & --- & $+3.3$ & --- & --- \\
            
            30 & H$_2$O\textsubscript{(grain)} & $<$ limit & $<$ limit & & --- & $<$ limit & --- & --- & --- & $-3.3$ & --- & --- \\
            
            30 & CO$_2$\textsubscript{(grain)} & $<$ limit & $<$ limit & & --- & $<$ limit & --- & $-5.4\times 10^1$ & --- & --- & --- & --- \\
            \hline
            \hline
            Radfield   &  &   &  & & &  &  &  &  &  &  & \\
            \hline
                    
            10 & H$_2$O & --- & $-3.8$ & & --- & $+3.9$ & --- & $+3.6$ & --- & $+6.1$ & --- & --- \\
            
            10 & CS & $-7.7$ & $-1.3\times 10^1$ & & --- & $+3.4\times 10^1$ & --- & $+1\times 10^1$ & --- & $+1.2\times 10^1$ & --- & $+7$ \\
            
            10 & NH$_3$\textsubscript{(grain)} & $<$ limit & $<$ limit & & $+3.1$ & --- & --- & --- & --- & --- & --- & --- \\
            
            10 & N$_2$H$^+$ & $<$ limit & $<$ limit & & --- & --- & --- & $+6.8$ & --- & --- & --- & --- \\
            
            10 & NH$_3$ & --- & $-8.3$ & & $+3.4$ & $+1.6\times 10^1$ & --- & --- & --- & $+5.7$ & --- & $+3$ \\
            
            10 & CO\textsubscript{(grain)} & $<$ limit & $<$ limit & & --- & $-6.1\times 10^6$ & --- & --- & --- & $+3.5$ & --- & --- \\
            
            10 & HCO$^+$ & --- & $-7.6$ & & --- & --- & --- & $+4.8$ & --- & $+8$ & --- & --- \\

            10 & CO$_2$\textsubscript{(grain)} &$<$ limit & $<$ limit & & --- & $-2.8\times 10^3$ & --- & $-6.2$ & --- & --- & --- & $-5.8$ \\
            
            100 & H$_2$O & --- & --- & & --- & $+2.8\times 10^1$ & --- & $+3.6$ & --- & $+6.1$ & --- & --- \\
            
            100 & CS & --- & $-7.4$ & & --- & $+1.7\times 10^1$ & --- & $+1\times 10^1$ & --- & $+1.2\times 10^1$ & --- & $+6.9$ \\
            
            100 & N$_2$H$^+$ & $<$ limit & $<$ limit & & $<$ limit &$<$ limit & --- & $+6.8$ & --- & --- & --- & --- \\
            
            100 & NH$_3$ & --- & --- & & $+3.8$ & $+7.2\times 10^3$ & --- & --- & --- & $+5.6$ & --- & $+3$ \\  
            
            100 & CO\textsubscript{(grain)} & $<$ limit & $<$ limit & & $<$ limit & $<$ limit & --- & --- & --- & $+3.5$ & --- & --- \\
            
            100 & HCO$^+$ & --- & --- & & --- & $+3.1\times 10^1$ & --- & $+4.8$ & --- & $+8$ & --- & --- \\
            
            100 & CO$_2$\textsubscript{(grain)} & $<$ limit & $<$ limit & & $<$ limit & $<$ limit & --- & $-6.2$ & --- & --- & --- & $-5.7$ \\
            \hline
            \hline
            Zeta  &  &   &  & & &  &  &  &  &  &  & \\
            \hline
                    
            10 & H$_2$O & $-4.7$ & $-4.1\times 10^1$ & & --- & $-7.8$ & --- & $+6.4$ & --- & --- & --- & --- \\
                    
            10 & CS & $-5.1\times 10^1$ & $<$ limit & & $+4.2$ & --- & --- & $+1.7\times 10^1$ & --- & $+4.8$ & --- & $+3.8$ \\
    
            10 & NH$_3$\textsubscript{(grain)} & $<$ limit & $<$ limit & & --- & $-7.4\times 10^5$ & --- & --- & --- & --- & --- & --- \\
                    
            10 & N$_2$H$^+$ & $<$ limit & $<$ limit & & --- & $-9.6$ & --- & --- & --- & $+4.9$ & --- & $+3.9$ \\
                    
            10 & NH$_3$ & $-7.1$ & $-1.1\times 10^2$ & & --- & $-1.2\times 10^1$ & --- & --- & --- & --- & --- & --- \\
                    
            10 & CO\textsubscript{(grain)} & $<$ limit & $<$ limit & & $-4.6\times 10^5$ & $-2.7\times 10^7$ & --- & $-3.8$ & --- & --- & --- & --- \\
                    
            10 & HCO$^+$ & $-9.7$ & $-6.7\times 10^1$ & & --- & $-6.3$ & --- & --- & --- & $+3.8$ & --- & $+3.5$ \\
                    
            10 & H$_2$O\textsubscript{(grain)} & $-5.6$ & $<$ limit & & --- & $-2.8\times 10^6$ & --- & --- & --- & --- & --- & --- \\
                    
            10 & CO$_2$\textsubscript{(grain)} & $<$ limit & $<$ limit & & $-1.5\times 10^2$ & $<$ limit & --- & $-6.6\times 10^1$ & --- & --- & --- & --- \\

            100 & H$_2$O & $-1.3\times 10^1$ & $-2.1\times 10^2$ & & $-3.1$ & $-4.6\times 10^1$ & --- & $-4.8$ & --- & $+3.6$ & --- & --- \\
            
            100 & CS & $<$ limit & $<$ limit & & --- & $-1\times 10^3$ & --- & --- & --- & $+6.2$ & --- & --- \\
            
            100 & NH$_3$\textsubscript{(grain)} & $<$ limit & $<$ limit & & $-4$ & $-1.8\times 10^1$ & --- & $<$ limit & --- & --- & --- & --- \\
            
            100 & N$_2$H$^+$ & $<$ limit & $<$ limit & & --- & $<$ limit & --- & $-1.1\times 10^1$ & --- & --- & --- & --- \\
            
            100 & NH$_3$ & $-4.4\times 10^1$ & $-2\times 10^1$ & & $-3.8$ & $-2.1\times 10^1$ & --- & $-8.6$ & --- & --- & --- & --- \\
            
            100 & CO\textsubscript{(grain)} & $<$ limit & $<$ limit & & --- & $<$ limit & $-3.2$ & $-3.9\times 10^6$ & --- & --- & --- & --- \\
            
            100 & HCO$^+$ & $<$ limit & $<$ limit & & --- & $-3.2\times 10^1$ & --- & $-8.6$ & --- & --- & --- & --- \\
           
            100 & H$_2$O\textsubscript{(grain)} & $<$ limit & $<$ limit & & $-3.9$ & $-1.4\times 10^1$ & --- & $-4.9\times 10^5$ & --- & --- & --- & --- \\
            
            100 & CO$_2$\textsubscript{(grain)} & $<$ limit & $<$ limit & & $<$ limit & $<$ limit & $-4.4$ & $<$ limit & --- & $-4.6$ & --- & --- \\
    
            \hline
        \end{tabular}

        \caption{Table showing results of the ionization rate dependency on abundances with varying parameters. Values shown are representing increase($+$) or decrease($-$) of abundances compared to the basic model and by what factor the abundances differed. This table only shows abundance changes greater than a factor of 3 and any value marked "$<$ limit" is below the lower abundance limit of 10$^{-13}$.}
        \label{tab:ionBasLargeResults}
    \end{sidewaystable*}

    \begin{table*}[ht]
        \centering
    
        \begin{tabular}{m{1.5cm} m{1cm} m{0.5cm}m{1.4cm}m{1.4cm}m{1cm}m{1cm}m{1cm}m{1cm}m{1cm}m{1cm}}
            \hline \hline
            \multicolumn{11}{c}{Post-Collapse} \\
            &&&&&&&&&& \\
            \multicolumn{11}{c}{Dissociation Dependency}\\
            \hline
            & Density: & & \multicolumn{2}{c}{10$^4$ cm$^{-3}$} & \multicolumn{2}{c}{10$^5$ cm$^{-3}$} & \multicolumn{2}{c}{10$^6$ cm$^{-3}$} & \multicolumn{2}{c}{10$^7$ cm$^{-3}$}\\
            
            Radfield  & Species & & L & H & L & H & L & H & L & H\\
            \hline

            100 & NH$_3$\textsubscript{(grain)}  & & $+4.6$ & --- & --- & --- & --- & --- & --- & --- \\
            
            100 & CO\textsubscript{(grain)} & & $+4.5$ & --- & --- & --- & --- & --- & --- & --- \\

            100 & H$_2$O\textsubscript{(grain)} & & $+4.5$ & --- & --- & --- & --- & --- & --- & --- \\
            
            100 & CO$_2$\textsubscript{(grain)} & & $+4.2$ & --- & --- & --- & --- & --- & --- & --- \\
            \hline 
            \hline
            Zeta  &&&&&&&&&& \\
            \hline

            10 & CO\textsubscript{(grain)}  & & $+1.4\times 10^{3}$ & --- & --- & --- & --- & --- & --- & --- \\
            
            \hline
            &&&&&&&&&& \\
            \multicolumn{11}{c}{Excited Species}\\
            \hline
            & Density: & & \multicolumn{2}{c}{10$^4$ cm$^{-3}$} & \multicolumn{2}{c}{10$^5$ cm$^{-3}$} & \multicolumn{2}{c}{10$^6$ cm$^{-3}$} & \multicolumn{2}{c}{10$^7$ cm$^{-3}$}\\
            
            \hline
            
            Temperature (K)  & & & &  &  &  \\
            \hline

            30 & N$_2$H$^+$  & & \multicolumn{2}{c}{---} & \multicolumn{2}{c}{$-3.9$} & \multicolumn{2}{c}{---} & \multicolumn{2}{c}{---} \\
            \hline 
            \hline
            Zeta  & & &  &  &  &  \\
            \hline
            
            10 & CO$_2$\textsubscript{(grain)}  & & \multicolumn{2}{c}{$-6.5$} &  \multicolumn{2}{c}{---} &  \multicolumn{2}{c}{---} &  \multicolumn{2}{c}{---} \\

            100 & H$_2$O  & & \multicolumn{2}{c}{---} & \multicolumn{2}{c}{$+3.7$} & \multicolumn{2}{c}{---} & \multicolumn{2}{c}{---} \\
            
            100 & CS  & & \multicolumn{2}{c}{---} & \multicolumn{2}{c}{$-7\times 10^1$} & \multicolumn{2}{c}{$-1.8\times 10^3$} & \multicolumn{2}{c}{$-7.1$} \\
            
            100 & N$_2$H$^+$ & & \multicolumn{2}{c}{---} & \multicolumn{2}{c}{---} & \multicolumn{2}{c}{$+3.5$} & \multicolumn{2}{c}{$+3.4$} \\
            
            100 & NH$_3$  & & \multicolumn{2}{c}{---} & \multicolumn{2}{c}{---} & \multicolumn{2}{c}{$+4.6$} & \multicolumn{2}{c}{$+3.9$} \\
            
            100 & CO\textsubscript{(grain)} & & \multicolumn{2}{c}{---} & \multicolumn{2}{c}{$-1.7\times 10^8$} & \multicolumn{2}{c}{$-8.1\times 10^9$} & \multicolumn{2}{c}{$-2.2\times 10^7$} \\
            
            100 & HCO$^+$ & & \multicolumn{2}{c}{---} & \multicolumn{2}{c}{$-2.3\times 10^2$} & \multicolumn{2}{c}{$-4.9\times 10^3$} & \multicolumn{2}{c}{$-2.9\times 10^1$} \\
            
            100 & H$_2$O\textsubscript{(grain)} & & \multicolumn{2}{c}{---} & \multicolumn{2}{c}{$-3.9$} & \multicolumn{2}{c}{---} & \multicolumn{2}{c}{---} \\
            
            100 & CO$_2$\textsubscript{(grain)} & & \multicolumn{2}{c}{---} & \multicolumn{2}{c}{$-2.2\times 10^2$} & \multicolumn{2}{c}{$-1\times 10^4$} & \multicolumn{2}{c}{$-3.8\times 10^{10}$} \\
    
            \hline

        \end{tabular}
        \caption{Table showing the only conditions where the inclusion of the dissociation rate dependency or the inclusion of excited species had any notable effect on abundances. Values shown are representing increase($+$) or decrease($-$) of abundances compared to the basic model and by what factor the abundances differed. This table only shows abundance changes greater than a factor of 3 and any value marked "$<$ limit" is below the lower abundance limit of 10$^{-13}$.}
        \label{tab:dissexcitedResults}
    \end{table*}

    \begin{table*}[ht]
        \centering
    
        \begin{tabular}{m{1.3cm} m{1cm} m{0.3cm}m{1cm}m{1cm}m{1.4cm}m{1.4cm}m{1.4cm}m{1.4cm}m{1.4cm}m{1.4cm}}
            \hline \hline
            & & & \multicolumn{7}{c}{Post-Collapse} \\
            \hline
            & Density: & & \multicolumn{2}{c}{10$^4$ cm$^{-3}$} & \multicolumn{2}{c}{10$^5$ cm$^{-3}$} & \multicolumn{2}{c}{10$^6$ cm$^{-3}$} & \multicolumn{2}{c}{10$^7$ cm$^{-3}$}\\
            \hline
            & Species    & & L & H & L & H & L & H & L & H\\
            \hline
 
            & CO\textsubscript{(grain)}  & & --- & $-3.3$ & --- & $-5.4$ & --- & $-6.1$ & --- & --- \\
                
            & HCO$^+$  & & --- & --- & --- & --- & --- & --- & $-3.4$ & --- \\
                
            & H$_2$O\textsubscript{(grain)}  & & --- & $-4.1$ & --- & --- & --- & --- & --- & --- \\
                
            & CO$_2$\textsubscript{(grain)}  & & --- & $-3.6$ & --- & $-6.7$ & $-5.9$ & $-3.7$ & $-5$ & $-7.4$ \\
                
            \hline
            \hline
            Temperature (K)   &  & &  &  &  &  &  &  & & \\
            \hline

            30 & N$_2$H$^+$  & & --- & --- & --- & --- & --- & --- & $<$ limit & $+5.8$ \\

            \hline
            \hline 
            Radfield  &  & &  &  &  &  &  &  & & \\
            \hline

            10 & CO\textsubscript{(grain)}  & & --- & --- & --- & $-5.3\times 10^1$ & --- & $-6$ & --- & --- \\
            
            10 & HCO$^+$  & & --- & --- & --- & --- & --- & --- & $-3.4$ & --- \\
            
            10 & H$_2$O\textsubscript{(grain)}  & & --- & $-4$ & --- & --- & --- & --- & --- & ---  \\
            
            10 & CO$_2$\textsubscript{(grain)}  & & --- & $-3.4$ & --- & $-6.7$ & $-5.9$ & $-3.7$ & $-4.8$ & $-7.4$ \\

            100 & CO\textsubscript{(grain)} & & $<$ limit & --- & --- & $-5.1\times 10^1$ & --- & $-6$ & --- & --- \\
            
            100 & HCO$^+$ & & --- & --- & --- & --- & --- & --- & $-3.4$ & --- \\

            100 & CO$_2$\textsubscript{(grain)} & & $<$ limit & $<$ limit & --- & $-6.8$ & $-5.9$ & $-3.7$ & $-4.9$ & $-7.4$ \\
    
            \hline
            \hline 
            Zeta  &  & &  &  &  &  &  &  & &\\
            \hline
            10 & H$_2$O  & & --- & --- & --- & $+3.5$ & --- & --- & --- & --- \\
            
            10 & CS  & & --- & --- & --- & $-2.4\times 10^2$ & --- & $-4.1\times 10^1$ & --- & --- \\
            
            10 & NH$_3$\textsubscript{(grain)}  & & --- &--- & --- & $-3.7$ & --- & --- & --- & --- \\
            
            10 & N$_2$H$^+$  & & --- & --- & --- & --- & --- & $+3.2$ & --- & --- \\
            
            10 & NH$_3$  & & --- & --- & --- & --- & --- & $+4.5$ & --- & --- \\
            
            10 & CO\textsubscript{(grain)}  & & --- & --- & $-3.7$ & $<$ limit & --- & $-2.4\times 10^8$ & --- & --- \\
            
            10 & HCO$^+$  & & --- & --- & --- & $-6.9\times 10^2$ & --- & $-1.2\times 10^2$ & --- & --- \\
            
            10 & H$_2$O\textsubscript{(grain)}  & & --- & --- & --- & $-8.6$ & --- & --- & --- & --- \\
            
            10 & CO$_2$\textsubscript{(grain)}  & & --- & $<$ limit & $-4.9$ & $-4.9\times 10^2$ & --- & $-2.6\times 10^2$ & --- & --- \\

            100 & H$_2$O  & & --- & --- & $+3.3$ & --- & --- & --- & --- & --- \\
            
            100 & CS  & & --- & --- & --- & --- & $-9\times 10^4$ & --- & --- & $-1.4\times 10^3$ \\
            
            100 & NH$_3$\textsubscript{(grain)}  & & --- & --- & $-4.8$ & --- & --- & --- & --- & $+4.4$ \\
            
            100 & N$_2$H$^+$  & & --- & $<$ limit & --- & --- & $+3.6$ & --- & --- & $+1.2\times 10^1$ \\
            
            100 & NH$_3$  & & --- & --- & --- & --- & $+4.5$ & $+3$ & $+3.3$ & $+9.1$ \\  
            
            100 & CO\textsubscript{(grain)} & & --- & $<$ limit & $-9\times 10^7$ & --- & $-3.2\times 10^1$ & --- & $-7.4\times 10^6$ & --- \\
            
            100 & HCO$^+$ & & --- & --- & $-5.9\times 10^2$ & --- & $-2.5\times 10^5$ & --- & $-9.8$ & $-5.8\times 10^3$ \\
            
            100 & H$_2$O\textsubscript{(grain)} & & --- & --- & $-1.2\times 10^1$ & --- & --- & $-3.7$ & --- & --- \\
            
            100 & CO$_2$\textsubscript{(grain)} & & $<$ limit & $<$ limit & $-4.9\times 10^2$ & --- & $-4.7\times 10^5$ & --- & $-1.3\times 10^1$ & $-8.9\times 10^3$ \\
    
            \hline
        \end{tabular}
        \caption{Table showing results of the combined dissociation rate dependency and excited species on abundances under standard conditions and with varying parameters. Values shown are representing increase($+$) or decrease($-$) of abundances compared to the basic model and by what factor the abundances differed. This table only shows abundance changes greater than a factor of 3 and any value marked "$<$ limit" is below the lower abundance limit of 10$^{-13}$.}
        \label{tab:combinedResults}
    \end{table*}

\end{document}